%% file: main.tex
\documentclass{llncs}

\usepackage{url}
\usepackage{amsmath}
\usepackage{amssymb}
\usepackage{times} 
\usepackage{paralist}
\usepackage{mathrsfs}
\usepackage{listings}
\usepackage{longtable}
\usepackage{tikz}
\usetikzlibrary{positioning,arrows.meta,automata}
\usepackage{pgfplots}
\usepgfplotslibrary{fillbetween}
\usepackage{graphicx}
\usepackage{subcaption} 
\usepackage{wrapfig}
\usepackage{multicol}
\usepackage{macros}
\usepackage{hyperref}
\usepackage[capitalise]{cleveref}
\usepackage[linecolor=red,bordercolor=red,backgroundcolor=white,textsize=scriptsize]{todonotes}
\usepackage[outdir=./fig/]{epstopdf}

\usepackage{algorithm}
\usepackage[noend]{algpseudocode}
\makeatletter

\let\c@table\c@figure

\begingroup \catcode `|=0 \catcode `[= 1
\catcode`]=2 \catcode `\{=12 \catcode `\}=12
\catcode`\\=12 |gdef|@xcomment#1\end{comment}[|end[comment]]
|endgroup

\def\@comment{\let\do\@makeother \dospecials\catcode`\^^M=10\def\par{}}

\def\begincomment{\@comment\@xcomment}

\makeatother

\begin{document}	


\title{Strategy Representation by Decision Trees\\ with Linear Classifiers\vspace{-2mm}}

\author{Pranav Ashok$^1$, Tom\'{a}\v{s} Br\'{a}zdil$^2$, Krishnendu Chatterjee$^3$,\\Jan K\v{r}et\'{i}nsk\'{y}$^1$, Christoph H. Lampert$^3$, and Viktor Toman$^3$}
\institute{$^1$ Technical University of Munich, Munich, Germany\\$^2$ Masaryk University, Brno, Czech Republic\\$^3$ IST Austria, Klosterneuburg, Austria}

\maketitle

\input{1_introduction}

\input{2_preliminaries}

\input{3_trees}

\input{4_classifiers}

\input{5_experiments}

\input{6_related}

\input{7_conclusion}


\bibliographystyle{abbrv}
\bibliography{0_literature}

\clearpage
\input{9_appendix}

\end{document}

%% file: 1_introduction.tex
\begin{abstract}
Graph games and Markov decision processes (MDPs) are standard models in 
reactive synthesis and verification of probabilistic systems with nondeterminism. 
The class of $\omega$-regular winning conditions; e.g., 
safety, reachability, liveness, parity conditions; 
provides a robust and expressive specification formalism for properties
that arise in analysis of reactive systems.
The resolutions of nondeterminism in games and MDPs are represented as strategies,
and we consider succinct representation of such strategies.
The decision-tree data structure from machine learning retains the flavor of decisions 
of strategies and allows entropy-based minimization to obtain succinct trees.
However, in contrast to traditional machine-learning problems where small errors
are allowed, for winning strategies in graph games and MDPs no error is allowed, 
and the decision tree must represent the entire strategy.
In this work we propose decision trees with linear classifiers for representation of strategies 
in graph games and MDPs. We have implemented strategy representation using
this data structure and we present experimental results for problems
on graph games and MDPs, which show that this new data structure presents
a much more efficient strategy representation as compared to standard decision trees.
\end{abstract}

\section{Introduction}\label{sec:intro}

\smallskip\noindent{\bf Graph games and MDPs.}
Graph games and Markov decision processes (MDPs) are classical 
models in reactive synthesis. 
In graph games, there is a finite-state graph, where the 
vertices are partitioned into states controlled by the two players, 
namely, player~1 and player~2, respectively. 
In each round the state changes according to a transition 
chosen by the player controlling the current state.
Thus, the outcome of the game being played for an infinite 
number of rounds, is an infinite path through the graph, which is 
called a play.
In MDPs, instead of an adversarial player~2, there are probabilistic choices. 
An objective specifies a subset of plays that are satisfactory.
A strategy for a player is a recipe to specify the choice of the 
transitions for states controlled by the player.
In games, given an objective, a winning strategy for a player from a state ensures 
the objective irrespective of the strategy of the opponent.
In MDPs, given an objective, an \emph{almost-sure winning strategy} from a state ensures
the objective with probability~1.

\smallskip\noindent{\bf Reactive synthesis and verification.}
The above models play a crucial role in various areas of computer science,
in particular analysis of reactive systems.
In reactive-system analysis, the vertices and edges of a graph 
represent the states and transitions of a reactive system, and the two 
players represent controllable versus uncontrollable decisions during the
execution of the system.
The reactive synthesis problem asks for construction of winning strategies
in adversarial environment, and almost-sure winning strategies in 
probabilistic environment.
The reactive synthesis for games has a long history, 
starting from the work of Church~\cite{Church62,Buchi62} and has been extensively 
studied~\cite{Rabin69,BuchiLandweber69,GH82,McNaughton93}, with many applications 
in synthesis of discrete-event and reactive systems~\cite{RamadgeWonham87,PnueliRosner89}, 
modeling~\cite{Dill89book,ALW89}, refinement~\cite{FairSimulation}, 
verification~\cite{DetectingErrorsBeforeReaching,AHK02}, testing~\cite{GurevichTest}, 
compatibility checking \cite{InterfaceAutomata}, etc.
Similarly, MDPs have been extensively used in verification of probabilistic
systems~\cite{POMC,PRISM,STORM}.
In all the above applications, the objectives are $\omega$-regular, 
and the $\omega$-regular sets of infinite paths provide an important and robust 
paradigm for reactive-system specifications~\cite{MannaPnueliVol1,Thomas97}.

\smallskip\noindent{\bf Strategy representation.}
The strategies are the most important objects as they represent the witness to winning/almost-sure winning.
The strategies can represent desired controllers in reactive synthesis and protocols,
and formally they can be interpreted as a lookup table that specifies for every controlled 
state of the player the transition to choose.
As a data structure to represent strategies, there are some desirable properties, which 
are as follows: (a)~{\em succinctness}, i.e., small strategies are desirable, since smaller 
strategies represent efficient controllers; 
(b)~{\em explanatory}, i.e., the representation explains the decisions of the strategies.
While one standard data structure representation for strategies is binary decision diagrams
(BDDs)~\cite{Akers78,Bryant86}, recent works have shown that decision trees~\cite{DBLP:journals/ml/Quinlan86,Mitchell1997} 
from machine learning provide an attractive alternative data structure for strategy
representation~\cite{DBLP:conf/cav/BrazdilCCFK15,DBLP:conf/tacas/BrazdilCKT18}. 
The two key advantages of decision trees are: 
(a)~Decision trees utilize various predicates to make decisions and thus retain the inherent flavor 
of the decisions of the strategies; and
(b)~there are entropy-based algorithmic approaches for decision tree minimization~\cite{DBLP:journals/ml/Quinlan86,Mitchell1997}.
However, one of the key challenges in using decision trees for strategy representation is 
that while in traditional machine-learning applications errors are allowed,
for winning and almost-sure winning strategies errors are not permitted.

\smallskip\noindent{\bf Our contributions.}
While decision trees are a basic data structure in machine learning, their various extensions
have been considered. In particular, they have been extended with linear
classifiers~\cite{breiman84,Quinlan92,frank1998using,landwehr2003logistic}.
Informally, a linear classifier is a predicate that checks inequality of a linear combination of
variables against a constant.
In this work, we consider decision trees with linear classifiers for strategy representation 
in graph games and MDPs, which has not been considered before.
First, for representing strategies where no errors are permitted, we present a method 
to avoid errors both in decision trees as well as in linear classification.
Second, we present a new method (that is not entropy-based) for choosing predicates
in the decision trees, which further improves the succinctness of decisions trees with linear classifiers.
We have implemented our approach, and applied it to examples of reactive synthesis from 
SYNTCOMP benchmarks~\cite{DBLP:journals/corr/JacobsBBKPRRSST16}, model-checking
examples from PRISM benchmarks~\cite{PRISMbenchmarks}, and synthesis of randomly generated LTL formulae~\cite{Pnueli77}.
Our experimental results show significant improvement in succinctness of strategy representation
with the new data structure as compared to standard decision trees.

%% file: 2_preliminaries.tex
\section{Stochastic Graph Games and Strategies}\label{sec:prelim}

\subsection{Informal description}\label{subsec:prelim_informal}

\smallskip\noindent{\bf Stochastic graph games.}
We denote the set of probability distributions over a finite set $X$ as $\dist(X)$.
A {\em stochastic graph game} is a tuple $G=\tuple{S_1, S_2, A_1, A_2, \trans}$, 
where:
\begin{compactenum}[--]
\item $S_1$ and $S_2$ is a finite set of states for player~1 and player~2, respectively, and
\mbox{$S = S_1 \cup S_2$} denotes the set of all states;
\item $A_1$ and $A_2$ is a finite set of actions for player~1 and player~2, respectively, and
\mbox{$A = A_1 \cup A_2$} denotes the set of all actions; and
\item $\trans \colon (S_1 \times A_1) \cup (S_2 \times A_2) \to \dist(S)$ is a transition function 
that given a player~1 state and a player~1 action, or a player~2 state and 
a player~2 action, gives the probability distribution over the successor states.
\end{compactenum}
We consider two special cases of stochastic graph games, namely:
\begin{compactenum}[--]
\item \emph{graph games}, where for each $(s,\!a)$ in the domain of $\trans$, $\trans(s,\!a)(s')\!=\!1$ for some $s' \in S$.
\item \emph{Markov decision processes (MDPs)}, where $S_2 = \emptyset$ and $A_2 = \emptyset$.
\end{compactenum}
We consider stochastic graph games with several classical objectives, namely,
\emph{safety} (resp. its dual \emph{reachability}),
\emph{B\"{u}chi} (resp. its dual \emph{co-B\"{u}chi}),
and parity objectives.

\smallskip\noindent{\bf Stochastic graph games with variables.}
Consider a finite subset of natural numbers $X \subseteq \Nats$, and a finite set $\Var$ of variables over $X$,
partitioned into state-variables and action-variables $\Var = \Var_S \uplus \Var_A$ ($\uplus$ denotes a disjoint union).
A \emph{valuation} is a function that assigns values from $X$ to the variables.
Let $X^{\Var_S}$ (resp., $X^{\Var_A}$) denote the set of all valuations to the state-variables (resp., the action-variables).
We associate a stochastic graph game $G = \tuple{S_1, S_2, A_1, A_2, \trans}$ with a set of variables $\Var$,
such that (i) each state $s \in S$ is associated with a unique valuation $\mathit{val}_s \in X^{\Var_S}$, and
(ii) each action $a \in A$ is associated with a unique valuation $\mathit{val}_a \in X^{\Var_A}$.

\input{fig/game_game}
\begin{example}\label{ex:game}
Consider a simple system that receives requests for two different channels A and B.
The requests become pending and at a later point a response handles a request for the
respective channel. A controller must ensure that (i) the request-pending queues do not
overflow (their sizes are 2 and 3 for channels A and B, respectively), and that
(ii) no response is issued for a channel without a pending request. The system
can be modeled by the graph game depicted in~\cref{fig:game_game}.
The states of player 1 (controller issuing responses) are labeled with valuations of state-variables
capturing the number of pending requests for channel A and B, respectively. For brevity of
presentation, the action labels (corresponding to valuations of a single action-variable)
are shown only outgoing from one state, with a straightforward generalization for all other
states of player 1. Further, for clarity of presentation, the labels of states and actions for player 2
(environment issuing requests, with filled blue-colored states and actions) are omitted.
The controller must ensure the safety objective of avoiding the four error states.
\end{example}

\smallskip\noindent{\bf Strategy representation.}
The algorithmic problem treated in this work considers representation of memoryless almost-sure
winning strategies for stochastic graph games with variables. Given a stochastic graph
game and an objective, a \emph{memoryless} strategy for player $i \in \{1, 2\}$ is
a function $\strat \colon S_i \to A_i$ that resolves the nondeterminism for player $i$
by choosing the next action based on the currently visited state. Further, a strategy is
\emph{almost-sure winning} if it ensures the given objective irrespective of the strategy
of the other player.
In synthesis and verification of reactive systems, the problems often reduce to
computation of memoryless almost-sure winning strategies for stochastic graph games,
where the state space and action space is represented by a set of variables.
In practice, such problems arise from various sources, e.g., AIGER specifications~\cite{AIGER},
LTL synthesis~\cite{Pnueli77}, PRISM model checking~\cite{PRISM}.

\smallskip
\input{2_prelim_detail}

%% file: fig/game_game.tex
\begin{figure}
\vspace{-0mm}
\begin{center}
\begin{tikzpicture}[draw=black, circle, semithick, ->, initial text = ]
\newcommand{\xs}{6mm}
\newcommand{\xss}{19mm}
\newcommand{\ys}{11mm}
\newcommand{\db}{blue!98!black}

	\node (00p1) [fill=\db] {};
	\node (00p2) [draw=black, left=\xs of 00p1, initial left] {$0,0$};
	\node (01p1) [fill=\db, right=\xss of 00p1] {};
	\node (01p2) [draw=black, left=\xs of 01p1] {$0,1$};	
	\node (02p1) [fill=\db, right=\xss of 01p1] {};
	\node (02p2) [draw=black, left=\xs of 02p1] {$0,2$};	
	\node (03p1) [fill=\db, right=\xss of 02p1] {};
	\node (03p2) [draw=black, left=\xs of 03p1] {$0,3$};	

	\node (10p1) [fill=\db, below=\ys of 00p1] {};
	\node (10p2) [draw=black, left=\xs of 10p1] {$1,0$};
	\node (11p1) [fill=\db, right=\xss of 10p1] {};
	\node (11p2) [draw=black, left=\xs of 11p1] {$1,1$};	
	\node (12p1) [fill=\db, right=\xss of 11p1] {};
	\node (12p2) [draw=black, left=\xs of 12p1] {$1,2$};	
	\node (13p1) [fill=\db, right=\xss of 12p1] {};
	\node (13p2) [draw=black, left=\xs of 13p1] {$1,3$};	

	\node (20p1) [fill=\db, below=\ys of 10p1] {};
	\node (20p2) [draw=black, left=\xs of 20p1] {$2,0$};
	\node (21p1) [fill=\db, right=\xss of 20p1] {};
	\node (21p2) [draw=black, left=\xs of 21p1] {$2,1$};	
	\node (22p1) [fill=\db, right=\xss of 21p1] {};
	\node (22p2) [draw=black, left=\xs of 22p1] {$2,2$};	
	\node (23p1) [fill=\db, right=\xss of 22p1] {};
	\node (23p2) [draw=black, left=\xs of 23p1] {$2,3$};	
	
	\node (e1) [rectangle, draw=black, right=\xs of 13p1, rounded corners=4pt] {ERROR};
	\node (e2) [rectangle, draw=black, below=2mm of 22p2, rounded corners=4pt] {ERROR};	
	\node (e3) [rectangle, draw=black, left=\xs of 10p2, rounded corners=4pt] {ERROR};
	\node (e4) [rectangle, draw=black, above=4mm of 01p1, rounded corners=4pt] {ERROR};

	\draw [] (00p2) -- (00p1);
	\draw [] (00p2) -- (e4);
	\draw [] (00p2) -- (e3);
	\draw [] (01p2) -- (01p1);
	\draw [] (01p2) -- (e4);
	\draw [] (01p2) -- (00p1);
	\draw [] (02p2) -- (02p1);
	\draw [] (02p2) -- (e4);
	\draw [] (02p2) -- (01p1);
	\draw [] (03p2) -- (03p1);
	\draw [] (03p2) -- (e4);
	\draw [] (03p2) -- (02p1);

	\draw [] (10p2) -- (10p1);
	\draw [] (10p2) -- (e3);
	\draw [] (10p2) -- (00p1);
	\draw [] (11p2) -- node [xshift=0mm, yshift=1.2mm] {w} (11p1);
	\draw [] (11p2) -- node [xshift=-3mm, yshift=-1mm] {rA} (01p1);
	\draw [] (11p2) -- node [xshift=0.4mm, yshift=1.4mm] {rB} (10p1);
	\draw [] (12p2) -- (12p1);
	\draw [] (12p2) -- (02p1);
	\draw [] (12p2) -- (11p1);
	\draw [] (13p2) -- (13p1);
	\draw [] (13p2) -- (03p1);
	\draw [] (13p2) -- (12p1);
	
	\draw [] (20p2) -- (20p1);
	\draw [] (20p2) -- (e3);
	\draw [] (20p2) -- (10p1);
	\draw [] (21p2) -- (21p1);
	\draw [] (21p2) -- (20p1);
	\draw [] (21p2) -- (11p1);
	\draw [] (22p2) -- (22p1);
	\draw [] (22p2) -- (21p1);
	\draw [] (22p2) -- (12p1);
	\draw [] (23p2) -- (23p1);
	\draw [] (23p2) -- (22p1);
	\draw [] (23p2) -- (13p1);
	
	\path (00p1) edge [bend left=20, color=\db] (00p2);
	\path (00p1) edge [bend right=20, color=\db] (01p2);	
	\path (00p1) edge [bend left=20, color=\db] (10p2);	
	\path (01p1) edge [bend left=20, color=\db] (01p2);
	\path (01p1) edge [bend right=20, color=\db] (02p2);	
	\path (01p1) edge [bend left=20, color=\db] (11p2);	
	\path (02p1) edge [bend left=20, color=\db] (02p2);
	\path (02p1) edge [bend right=20, color=\db] (03p2);	
	\path (02p1) edge [bend left=20, color=\db] (12p2);	
	\path (03p1) edge [bend left=20, color=\db] (03p2);
	\path (03p1) edge [bend right=00, color=\db] (e1);	
	\path (03p1) edge [bend left=20, color=\db] (13p2);	
	
	\path (10p1) edge [bend left=20, color=\db] (10p2);
	\path (10p1) edge [bend right=20, color=\db] (11p2);	
	\path (10p1) edge [bend left=20, color=\db] (20p2);	
	\path (11p1) edge [bend left=20, color=\db] (11p2);
	\path (11p1) edge [bend right=20, color=\db] (12p2);	
	\path (11p1) edge [bend left=20, color=\db] (21p2);	
	\path (12p1) edge [bend left=20, color=\db] (12p2);
	\path (12p1) edge [bend right=20, color=\db] (13p2);	
	\path (12p1) edge [bend left=20, color=\db] (22p2);	
	\path (13p1) edge [bend left=20, color=\db] (13p2);
	\path (13p1) edge [bend right=00, color=\db] (e1);	
	\path (13p1) edge [bend left=20, color=\db] (23p2);		
	
	\path (20p1) edge [bend left=20, color=\db] (20p2);
	\path (20p1) edge [bend right=20, color=\db] (21p2);	
	\path (20p1) edge [bend right=20, color=\db] (e2);	
	\path (21p1) edge [bend left=20, color=\db] (21p2);
	\path (21p1) edge [bend right=20, color=\db] (22p2);	
	\path (21p1) edge [bend left=00, color=\db] (e2);	
	\path (22p1) edge [bend left=20, color=\db] (22p2);
	\path (22p1) edge [bend right=20, color=\db] (23p2);	
	\path (22p1) edge [bend left=00, color=\db] (e2);	
	\path (23p1) edge [bend left=20, color=\db] (23p2);
	\path (23p1) edge [bend right=00, color=\db] (e1);	
	\path (23p1) edge [bend left=20, color=\db] (e2);	
		

\end{tikzpicture}
	\caption{A reactive system with two request channels.}
	\label{fig:game_game}
\end{center}
\vspace{-8mm}
\end{figure}
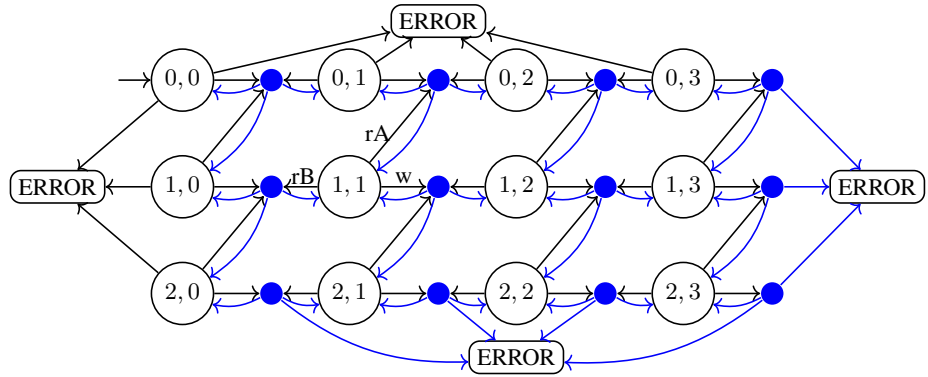

%% file: 2_prelim_detail.tex
\subsection{Detailed description}\label{subsec:prelim_detailed}

\smallskip\noindent{\bf Plays.}
Given a stochastic graph game $G$, a \emph{play} is an infinite sequence of
state-action pairs $\tuple{s_0 a_0 s_1 a_1 \ldots}$ such that for all $j \geq 0$ we have that
$(s_j, a_j) \in S_i \times A_i$ for some $i \in \set{1,2}$, and $(\trans(s_j, a_j)) (s_{j+1}) > 0$.
We denote by $\Play(G)$ the set of all plays in $G$.

\smallskip\noindent{\bf Objectives.}
An \emph{objective} for a stochastic graph game $G$ is a Borel set $\obj \subseteq \Play(G)$.
We consider the following objectives:
\begin{compactenum}[--]
\item \emph{Reachability and safety.}
Given a set $T \subseteq S$ of target states, the \emph{reachability} objective requires
that a state in $T$ is eventually visited.
Formally,
$\Reach(T) = \set{\tuple{s_0 a_0 s_1 a_1 \ldots} \in \Play(G) \mid \exists i : s_i \in T}$.
The dual of reachability objectives are \emph{safety} objectives,
where a set $F \subseteq S$ of safe states is given, and the safety objective requires
that only states in $F$ are visited.
Formally,
$\Safe(F) = \set{\tuple{s_0 a_0 s_1 a_1 \ldots} \in \Play(G) \mid \forall i : s_i \in F}$.
\item \emph{Parity.}
For an infinite play $\play$ we denote by $\Inf(\play)$ the set of states that
occur infinitely often in $\play$.
Let $p \colon S \to \Nats$ be a given \emph{priority function}.
The \emph{parity} objective $\Parity(p) = \{\play \in \Play(G) \mid \min\{p(s)
\mid s \in \Inf(\play)\} \text{ is even }\}$ requires that the minimum of the
priorities of the states visited infinitely often is even. The dual of the parity
objective requires that the minimum of the priorities visited infinitely often
is odd. For a special case of priority functions $p \colon S \to \set{0, 1}$,
the corresponding parity objective (resp., its dual) is
called \emph{B\"{u}chi} (resp., \emph{co-B\"{u}chi}).
\end{compactenum}

\smallskip\noindent{\bf Memoryless strategies.}
Given a stochastic graph game $G$, a strategy is a recipe for a player how to choose actions to
extend finite prefixes of plays. Specifically, a \emph{memoryless strategy} is a strategy where the
player performs each choice based solely on the currently visited state.
Formally, a memoryless strategy $\strat_1$ for player~1 is a function 
$\strat_1 \colon S_1 \to A_1$ that given a currently visited state chooses the next action.
Analogously, a memoryless strategy for player~2 is a function $\strat_2  \colon S_2 \to A_2$.
We denote by $\Strat_1(G)$ and $\Strat_2(G)$ the sets of all memoryless strategies for player~1 and 
player~2 in $G$, respectively.
Given two strategies $\strat_1 \in \Strat_1(G)$, $\strat_2 \in \Strat_2(G)$, and a starting 
state $s \in S$, they induce a unique probability measure
$\Prob_{s}^{\strat_1, \strat_2}$ over the Borel sets of $\Play(G)$.
In the special case of graph games, the two strategies and the starting state
induce a unique play $\play_{s}^{\strat_1, \strat_2} = \tuple{s_0 a_0 s_1 a_1 \ldots}$
such that $s_0=s$ and for all $j \geq 0$,
$s_j \in S_i$ and $a_j = \strat_i(s_j)$ for some $i \in \set{1,2}$.
The strategies we consider in this work are all memoryless strategies.

\smallskip\noindent{\bf Winning and almost-sure winning strategies.}
Given a stochastic graph game $G$ and an objective $\obj$, an \emph{almost-sure winning} strategy
$\strat_1 \in \Strat_1(G)$ from state $s \in S$ is a strategy such that for all 
strategies $\strat_2 \in \Strat_2(G)$ we have $\Prob_{s}^{\strat_1, \strat_2} (\obj) = 1$.
A fundamental result for stochastic graph games with parity (resp., safety/reachability)
objectives shows that (i) there is a memoryless almost-sure winning strategy
if and only if there is a general (i.e., utilizing the past and nondeterminism)
almost-sure winning strategy, and (ii) a memoryless almost-sure winning strategy satisfies the objective
with probability 1 even against general strategies of the opposing player~\cite{KrishPhD}.
In the special case of graph games, an almost-sure winning strategy $\strat_1 \in \Strat_1(G)$ ensures
for all $\strat_2 \in \Strat_2(G)$ that $\play_{s}^{\strat_1, \strat_2} \in \obj$, and is referred to as
\emph{winning} strategy.

\smallskip\noindent{\bf Reactive synthesis and strategies.}
In the analysis of reactive systems, most properties that arise in practice are 
$\omega$-regular objectives, which capture important desirable properties,
such as safety, liveness, fairness. The class of $\omega$-regular objectives is expressible
by the linear-time temporal logic (LTL) framework.
The problem of synthesis from LTL specifications has
received huge attention~\cite{ModelCheckBook}, and the LTL synthesis problem 
can be reduced to solving graph games with parity objectives.
Moreover, given a model and a specification, the fundamental model checking
problem asks to produce a witness that the model satisfies the specification.
In model checking of probabilistic systems, the witness for a property is a
policy that ensures the property almost-surely.
In such settings, it is natural to consider graph games and MDPs
where the state space and action space is represented by a set of variables.

%% file: 3_trees.tex
\vspace{-2mm}
\section{Decision Trees and Decision Tree Learning}\label{sec:trees}

Here we recall decision trees (DT), representing strategies by DT, and learning DT.


\input{fig/tree_example}

\smallskip\noindent{\bf Decision tree (DT)} over $\Nats^d$ is a tuple $\mathcal{T}=(T,\rho,\theta)$ where $T$ is a finite rooted binary (ordered) tree, 
$\rho$ assigns to every inner node an (in)equality predicate comparing arithmetical expressions over variables $\{x_1,\ldots,x_d\}$, and $\theta$
assigns to every leaf a value $\yes$ or $\no$.
The language $\mathcal{L}(\mathcal{T})\subseteq \Nats^d$ of the tree is defined as follows. 
For a vector $\vec{x}=(x_1,\ldots, x_d)\in \Nats^d$, we find a path $p$ from the root to a leaf such that for each inner node $n$ on the path,
$\rho(n)(\vec{x})=\true$ (i.e., the predicate $\rho(n)$ is satisfied with valuation $\vec{x}$) iff the~first child of $n$ is on $p$. 
Denote the leaf on this particular path by $\ell$. 
Then $\vec{x}$ is in the language $\mathcal{L}(\mathcal{T})$ of $\mathcal T$ iff $\theta(\ell)=\yes$.
Intuitively, $\mathcal{L}(\mathcal{T})$ captures the set of vectors \emph{accepted} by the tree $\mathcal{T}$,
i.e., vectors with accepting path in the tree (ending with $\yes$).
An example is illustrated in Fig.~\ref{fig:tree_example} with the first children connected with unbroken arrows and the second children with dashed ones.


The (usually finite) set of predicates in the co-domain of $\rho$ is denoted by $\preds$. 
In the example above $\preds$ are comparisons of variables to constants.


\vspace{-1mm}
\smallskip\noindent{\bf Representing strategies by DT} has been introduced in \cite{DBLP:conf/cav/BrazdilCCFK15}.
The dimension of data points here is $d = |\Var|$. The data points are natural tuples representing state-action pairs, thus we also write them as $(s,a)$.
The strategy induced by a decision tree $\mathcal T$ allows to play $a$ in $s$ iff $(s,a)\in \mathcal{L}(\mathcal{T})$.

A given input strategy $\strat \colon S_i \to A_i$ for player $i \in \{1,2\}$ defines the sets (i) $\good = \{\tuple{s,\strat(s)}\in S_i\times A_i\}$,
(ii) $\bad = \{\tuple{s,a} \in S_i\times A_i \mid a\not = \strat(s)\}$,
and (iii) $\train = \good \uplus \bad$ ($\uplus$ denotes a disjoint union).
Further, given a subset $\da \subseteq \train$, we define $\maxc(\da)$ as (i) $\yes$ if $|\da\cap\good|\geq |\da\cap\bad|$, and (ii) $\no$ otherwise.
When strategies need to be represented exactly, as in the case of games, the trees have to classify all decisions correctly \cite{DBLP:conf/tacas/BrazdilCKT18}.
This in turn causes difficulties not faced in standard DT learning \cite{Mitchell1997}, as described below.


\begin{example}\label{ex:strat}
Consider the reactive system and the corresponding game described in~\cref{ex:game}.
Consider a strategy $\strat$ for the controller (player 1) in this system that (i) waits in state $(0,0)$,
(ii) issues a response for channel B when there are more pending requests for channel B
than pending requests for channel A, and (iii) issues a response for channel A in all other cases. Then, the
strategy $\strat$ induces: $\good = \{ (0,0,w), (0,1,rB), (0,2,rB), (0,3,rB), (1,0,rA),
(1,1,rA), (1,2,rB), (1,3,rB), $\\$(2,0,rA), (2,1,rA), (2,2,rA), (2,3,rB)\}$, and
$\bad = \{\mathit{(p_A, p_B, act)} \in \{0,1,2\} \times \{0,1,2,3\} \times \mathit{\{w,rA,rB\}} \;|\; (p_A, p_B, act) \not \in \good \}$.
The task is to represent $\strat$ exactly, i.e., to accept all $\good$ examples and reject all $\bad$ examples.
\end{example}

\vspace{-1mm}
\smallskip\noindent{\bf Learning DT} from the set $\good$ of positive examples and the set $\bad$ of negative examples is described in \cref{alg:dt_basic}.
A node with all the data points is gradually split into offsprings until the point where each leaf contains only elements of $\good$ or only $\bad$.
Note that in the classical DT learning algorithms such as ID3~\cite{DBLP:journals/ml/Quinlan86}, one can also stop this process earlier
to prevent overfitting, which induces smaller trees with a classification error, unacceptable in the strategy representation.

\input{algo/dt_basic}
\vspace{-1mm}
\input{algo/split_ig}

The choice of the predicate to split a node with is described in \cref{alg:split_ig}. From the finite set
$\preds$\footnote{The set of considered predicates $\preds$ is typically domain-specific, and finitely restricted in a natural way.
In this work, we consider (in)equality predicates that compare values of variables to constants.
A natural finite restriction is to consider only constants that appear in the dataset.}
we pick the one which maximizes \emph{information gain} (i.e., decrease of entropy~\cite{Mitchell1997}).
Again, due to the need of fully expanded trees with no error, we need to guarantee that we can split all nodes with mixed
data even if none of the predicates provides any information gain in one step.
This issue is addressed in \cite{DBLP:conf/tacas/BrazdilCKT18} as follows. 
Whenever no positive information gain can be achieved by any predicate, a predicate is chosen according to a very simple
different formula using a heuristic that always returns a positive number.
One possible option suggested in \cite{DBLP:conf/tacas/BrazdilCKT18} is captured on Line~\ref{l:heur}.

%% file: fig/tree_example.tex
\begin{wrapfigure}[8]{r}{0.38\textwidth}
\vspace{-6mm}
\begin{center}
\begin{tikzpicture}[node distance = 1.2cm]

	\node (a1) [draw,rectangle,rounded corners=2pt] {$x_1 < 4$} ;
	\node (y1) [below left of=a1,draw,rectangle,rounded corners=2pt] {$\yes$};
	\node (a2) [below right of=a1,draw,rectangle,rounded corners=2pt]  {$x_1 = 7$};
	\node (y2) [below left of=a2,draw,rectangle,rounded corners=2pt] {$\yes$};
	\node (n1) [below right of=a2,draw,rectangle,rounded corners=2pt] {$\no$};

	\draw [->] (a1) -- node [xshift=-0.25cm, yshift=0.1cm] {} (y1);
	\draw [->, dashed] (a1) -- node [xshift=0.25cm, yshift=0.1cm] {} (a2);
	\draw [->] (a2) -- node [xshift=0.25cm, yshift=0.1cm] {} (y2);
	\draw [->, dashed] (a2) -- node [xshift=-0.25cm, yshift=0.1cm] {} (n1);

\end{tikzpicture}
	\caption{A decision tree for $\set{0,1,2,3,7} \subseteq \Nats^1$.}
	\label{fig:tree_example}
\end{center}
\end{wrapfigure}

%% file: algo/dt_basic.tex
{
\newcommand{\sat}{\mathit{sat}}
\newcommand{\uns}{\mathit{unsat}}
\begin{algorithm}
	\caption{Basic decision-tree learning algorithm\label{alg:dt_basic}}
	\begin{algorithmic}[1]
		\Statex \textbf{Input:} $\train \subseteq \Nats^{|\Var|}$ partitioned into subsets $\good$ and $\bad$.
		\Statex \textbf{Output:} A decision tree $\mathcal{T}$ such that $\mathcal{L}(\mathcal{T})\cap\train=\good$.
		\Statex /* train $\mathcal{T}$ on positive set $\good$ and negative set $\bad$ */
		\State $\mathcal{T} \gets (T= \set{\mathit{root}}, \rho = \emptyset, \theta = \emptyset)$
		\State $\mathsf{q} \gets \set{(\mathit{root}, \train)}$
		\While{$\mathsf{q}$ nonempty}
			\State $(\ell, \da_\ell) \gets \mathit{pop}_{\mathsf{q}}$
			\If{$\da_\ell \subseteq \good$ or $\da_\ell \subseteq \bad$}
				\State $\theta(\ell) \gets \maxc(\da_\ell)$				
			\Else
				\State $\rho(\ell) \gets$ predicate selected by a split procedure $\mathit{Split}(\da_\ell)$
				\State create children $\ell_\sat$ and $\ell_\uns$ of $\ell$
				\State $\mathit{push}_\mathsf{q}((\ell_\sat, \da_\ell[\rho(\ell)]))$, \,$\mathit{push}_\mathsf{q}((\ell_\uns, \da_\ell[\neg\rho(\ell)]))$
			\EndIf
		\EndWhile
		\State\Return $\mathcal{T}$
	\end{algorithmic}
\end{algorithm}
}

%% file: algo/split_ig.tex
{
\newcommand{\ig}{\mathsf{ig}}
\newcommand{\commentindent}{.48}
\renewcommand{\algorithmiccomment}[1]{\bgroup\hfill\makebox[\commentindent\textwidth][l]{$\triangleright$~#1}\egroup}
\begin{algorithm}
	\caption{Split procedure -- information gain\label{alg:split_ig}}
	\begin{algorithmic}[1]
		\Statex \textbf{Input:} $\da \subseteq \Nats^{|\Var|}$ partitioned into subsets $\da_G$ and $\da_B$.
		\Statex \textbf{Output:} A predicate $\pred$ maximizing information gain on $\da$.
		\State $\ig \gets \emptyset$
		\For{$\pred \in \preds$}
			\State $\ig(\pred) \gets$ information gain$(\da, \pred)$
		\EndFor
		\If{$\max_{\pred} \set{\ig(\pred)} = 0$}\Comment{condition checks if information gain failed}
			\For{$\pred \in \preds$}
				\State $\ig(\pred) \gets
\max \left\{\!
\frac{|\da_B[\neg\pred]|}{|\da[\neg\pred]|} \!+\! \frac{|\da_G[\pred]|}{|\da[\pred]|},
\frac{|\da_G[\neg\pred]|}{|\da[\neg\pred]|} \!+\! \frac{|\da_B[\pred]|}{\da[\pred]|} \label{l:heur}
\!\right\}$
			\EndFor
	
		\EndIf
		\State\Return $\argmax_{\pred} \set{\ig(\pred)}$
	\end{algorithmic}
\end{algorithm}
}

%% file: 4_classifiers.tex
\section{Decision Trees with Linear Classifiers}\label{sec:classifiers}

In this section, we develop an algorithm for constructing decision trees
with linear classifiers in the leaf nodes. As we are interested in representation
of winning and almost-sure winning strategies, we have to address the challenge
of allowing no error in the strategy representation. Thus we consider an algorithm
that provably represents a given strategy in its entirety. Furthermore, we present
a split procedure for decision-tree algorithms, which aims to propose predicates
leading into small trees with linear classifiers.

\subsection{Linear classifiers in the leaf nodes}

\input{fig/separability}
During the construction of a decision tree for a given dataset, each node
corresponds to a certain subset of the dataset. This subset exactly captures
the data points from the dataset that would reach the node starting from
the root and progressing based on the predicates visited along the travelled path
(as explained in~\cref{sec:trees}).
Notably, there might be other data points also reaching this node from the root,
however, they are not part of the dataset, and thus their outcome on the tree is
irrelevant for the correct dataset representation. This insight allows us to propose
a decision-tree algorithm with more expressive terminal (i.e., leaf) nodes,
and in this work we consider linear classifiers as the leaf nodes.

Given two vectors $\vec{a},\vec{b} \in \Reals^d$, their dot product (or scalar product) is defined as
$\vec{a} \dotproduct \vec{b} = \sum_{i=1}^d a_ib_i$.
Given a weight vector $\vec{w} \in \Reals^d$ and a bias term $b \in \Reals$,
a \emph{linear classifier} $c_{\vec{w},b} \colon \Reals^d \to {\yes, \no}$ is defined as
\vspace{-2mm}
\[
c_{\vec{w},b} (\vec{x}) =
\begin{cases}
	\yes & \vec{w} \dotproduct \vec{x} \geq b\\
	\no & \text{otherwise.}
\end{cases}
\vspace{-2mm}
\]
Informally, a linear classifier checks whether a linear combination of vector
values is greater than or equal to a constant.
Intuitively, we consider strategies as good and bad vectors of natural numbers, and
we use linear classifiers to decide for a given vector whether it is good
or bad. On a more general level, a linear classifier partitions the space $\Reals^d$
into two half-spaces, and a given vector gets classified based on the half-space it
belongs to.

Consider a finite dataset $\train \subseteq \Nats^d$ partitioned into subsets $\good$ and $\bad$.
A linear classifier
$c_{\vec{w},b}$ \emph{separates} $\train$, if for every $\vec{x} \in \train$
we have that \mbox{$c_{\vec{w},b}(\vec{x}) = \yes$} iff $\vec{x} \in \good$.
The corresponding decision problem asks, given a dataset $\train \subseteq \Nats^d$,
for existence of a weight vector $\vec{w} \in \Reals^d$ and bias $b \in \Reals$ such that
the linear classifier $c_{\vec{w},b}$ separates $\train$. In such a case we
say that $\train$ is linearly separable. \cref{fig:separability} provides an illustration.
There are efficient oracles for the decision problem
of linear separability, e.g., linear-programming solvers.

\input{fig/game_tree}
\begin{example}\label{ex:dt_tree}
We illustrate the idea of representing strategies by decision trees with linear
classifiers. Consider the game described in~\cref{ex:game} and the controller strategy
$\pi$ for this game described in~\cref{ex:strat}.
An example of a decision tree that represents the strategy $\strat$ is displayed in~\cref{fig:game_tree}.
The input samples with action $\mathit{w\:(wait)}$ end in and get classified by the leftmost linear classifier, and the samples
with action $\mathit{rB\:(responseB)}$ get classified by the rightmost linear classifier. Finally, the samples with
action $\mathit{rA\:(responseA)}$ are rejected if there are no pending requests to channel A, and otherwise
they get classified by the bottommost linear classifier.
Note that the decision tree accepts each sample from $\good$ and rejects each sample from $\bad$,
and thus indeed represents the strategy $\strat$.
\end{example}

\input{algo/dt_linc}

We are now ready to describe our algorithm for representing strategies as decision trees
with linear classifiers. \cref{alg:dt_linc} presents the pseudocode.
At the beginning, in Line~\ref{line:dt_linc_qinit} the queue is initiated with the root node and
the whole training set $\train$. Intuitively, the queue maintains the tree nodes
that are to be processed, and in every iteration of the loop (Line~\ref{line:dt_linc_loop})
one node $\ell$ gets processed. First, in Line~\ref{line:dt_linc_pop} the node $\ell$ gets popped
together with $\da_\ell$, which is the subset of $\train$ that would reach $\ell$ from the root node.
If $\da_\ell$ contains only samples from $\good$ (resp., only samples from $\bad$), then $\ell$
becomes a leaf node with $\yes$ (resp., $\no$) as the answer (Line~\ref{line:dt_linc_pure}).
If $\da_\ell$ contains samples from both, but is linearly separable by some classifier, then $\ell$
becomes a leaf node with this classifier (Line~\ref{line:dt_linc_classifier}).
Otherwise, $\ell$ becomes an inner node. In Line~\ref{line:dt_linc_pred} it gets assigned
a predicate by an external split procedure and in Line~\ref{line:dt_linc_children} two children of $\ell$
are created. Finally, in Line~\ref{line:dt_linc_push}, $\da_\ell$ is partitioned into the subset that
satisfies the chosen predicate of $\ell$ and the subset that does not, and the two children of $\ell$
are pushed into the queue with the two subsets, to be processed in later iterations.
Once there are no more nodes to be processed, the final decision tree is returned.
\input{4_example}
\input{4_proof}


\subsection{Splitting criterion for small decision trees with classifiers}

During construction of decision trees, the predicates for the inner nodes
are chosen based on a supplied metric, which heuristicly attempts to
select predicates leading into small trees. The entropy-based
\emph{information gain} is the most prevalent metric to construct decision trees,
in machine learning~\cite{Mitchell1997,DBLP:journals/ml/Quinlan86}
as well as in formal
methods~\cite{AlurBDF0JKMMRSSSSTU15,DBLP:conf/cav/BrazdilCCFK15,0001NMR16,DBLP:conf/tacas/NeiderSM16}. 
\cref{alg:split_ig} presents a split procedure utilizing information gain, supplemented
with a stand-in metric proposed in~\cite{DBLP:conf/tacas/BrazdilCKT18}.

In this section, we propose a new metric and we develop a split procedure
around it. When selecting predicates for the inner nodes, we exploit
the knowledge that in the descendants the data will be tested for linear
separability. Thus for a given predicate, the metric tries to estimate, roughly speaking,
how well-separable the corresponding data partitions are.
While the metric is well-studied in machine learning, to the best of our knowledge,
the corresponding decision-tree-split procedure is novel,
both in machine learning and in formal methods.

\input{fig/truefalse_posneg}

\smallskip\noindent{\bf True/false positive/negative.}
Consider a fixed linear classifier $c$, and a sample $\vec{x} \in \train$ such that
$c(\vec{x}) = \yes$. If $\vec{x} \in \good$, then $\vec{x}$ is a \emph{true positive} ($\tp$)
w.r.t. the classifier $c$, otherwise $\vec{x} \in \bad$ and thus $\vec{x}$ is a \emph{false positive} ($\fp$).
Consider a different sample $\vec{\bar{x}} \in \train$ such that $c(\vec{\bar{x}}) = \no$. If
$\vec{\bar{x}} \in \bad$, then $\vec{\bar{x}}$ is a \emph{true negative} ($\tn$), otherwise
$\vec{\bar{x}} \in \good$ and $\vec{\bar{x}}$ is a \emph{false negative} ($\fn$).
\cref{fig:truefalse_posneg} summarizes the terminology.

\smallskip\noindent{\bf True/false positive rate.}
Consider a fixed linear classifier $c$ and a fixed dataset $\train = \good \uplus \bad$.
We denote by $|\tp|$ the number of true positives within $\train$ w.r.t. the classifier $c$.
Similarly we denote $|\fp|$ for false positives. Then, the \emph{true positive rate} ($\tpr$) is
defined as $|\tp|/|\good|$, and the \emph{false positive rate} ($\fpr$) is $|\fp|/|\bad|$.
Intuitively, $\tpr$ describes the fraction of good samples that are correctly classified,
whereas $\fpr$ describes the fraction of bad samples that are misclassified as good.

\smallskip\noindent{\bf Area under the curve.}
Consider a fixed dataset $\train = \good \uplus \bad$ and a fixed weight vector $\vec{w} \in \Reals^d$.
In what follows we describe a metric that evaluates $\vec{w}$ w.r.t. $\train$.
First, consider a set of \emph{boundaries}, which are the dot products of $\vec{w}$ with samples from $\train$.
Formally, $\mathit{bnd} = \set{ \vec{w} \dotproduct \vec{x} \;|\; \vec{x} \in \train }$. Further, consider
$b_\mathit{none} = \max{\mathit{bnd}} + \varepsilon$ for some $\varepsilon > 0$.
Then, consider the set of linear classifiers that ``hit'' the boundaries, plus a classifier that rejects all samples.
Formally, $\mathit{cl} = \set{ c_{\vec{w}, b} \;|\; b \in \mathit{bnd} \cup \set{b_\mathit{none}}}$.
Now, the \emph{receiver operating characteristic} (ROC) is a curve that plots $\tpr$ against $\fpr$ for the classifiers in $\mathit{cl}$.
Intuitively, the ROC curve captures, for a fixed set of weights, how changing the bias term affects $\tpr$ and $\fpr$ of the resulting
classifier. Ideally, we want the $\tpr$ to increase rapidly when bias is weakened, while the $\fpr$ increases as little as possible.
We consider the area under the ROC curve (denoted $\auc \in [0,1]$) as the metric to evaluate
the weight vector $\vec{w}$ w.r.t. the dataset $\train$. Intuitively, the faster the $\tpr$ increases, and the slower the $\fpr$
increases, the bigger the area under the ROC curve ($\auc$) will be.

\cref{fig:roc} provides an intuitive illustration of the concept, where the weight vector is fixed as $\vec{w} = (1,0)$.
The classifiers $\mathit{cl}$ are then shown on the left subfigure, and the corresponding ROC curve
(with the shaded area under the curve -- $\auc$) is shown on the right subfigure. Note that the points in the ROC curve correspond
to the classifiers from $\mathit{cl}$, and they capture their ($\fpr$,$\tpr$). The extra point $(0/2,0/3)$ corresponds to
the classifier that rejects all samples.

\input{fig/roc}

\input{algo/split_auc}

\cref{alg:split_auc} presents a split procedure that uses $\auc$ as the metric to select predicates.
Each considered predicate partitions input $\da$ into the subset that satisfies the predicate and the subset that does not.
Then, in Lines~\ref{line:split_auc_sat}~and~\ref{line:split_auc_unsat}, two weight vectors are obtained by solving
the linear least squares problem on the data partitions. This is a classical problem in statistics with a known
closed-form solution, and~\cref{app:linls} provides detailed description of the problem. 
Finally, the score for the predicate
equals the sum of $\auc$ for the two weight vectors with respect to their corresponding data partitions (Line~\ref{line:split_auc_score}).
At the end, in Line~\ref{line:split_auc_ret} the predicate with maximum score is selected.

The choice of $\auc$ as the split metric is motivated by heuristicly estimating well-separability of data in the setting of strategy representation.
A simpler metric of \emph{accuracy} (i.e., the fraction of correctly classified samples) may seem as a natural choice for the estimate
of well-separability. However, in strategy representation, the data is typically very inbalanced, i.e., the sizes of $\good$ are typically much
smaller than the sizes of $\bad$. As a result, for all considered predicates the corresponding proposed classifiers focus heavily on
the $\bad$ samples and neglect the few $\good$ samples. Thus all classifiers achieve remarkable accuracy, which gives us little information
on the choice of a predicate. This is a well-known insight, as in machine learning, the accuracy metric is notoriously problematic
in the case of disproportionate classes. On the other hand, the $\auc$ metric, utilizing the invariance of bias, is able to focus also
on the sparse $\good$ subset, thus providing better estimates on well-separability.

%% file: fig/separability.tex
\begin{wrapfigure}[15]{r}{0.34\textwidth}
\vspace{-6mm}
\centering
\includegraphics[scale=0.16]{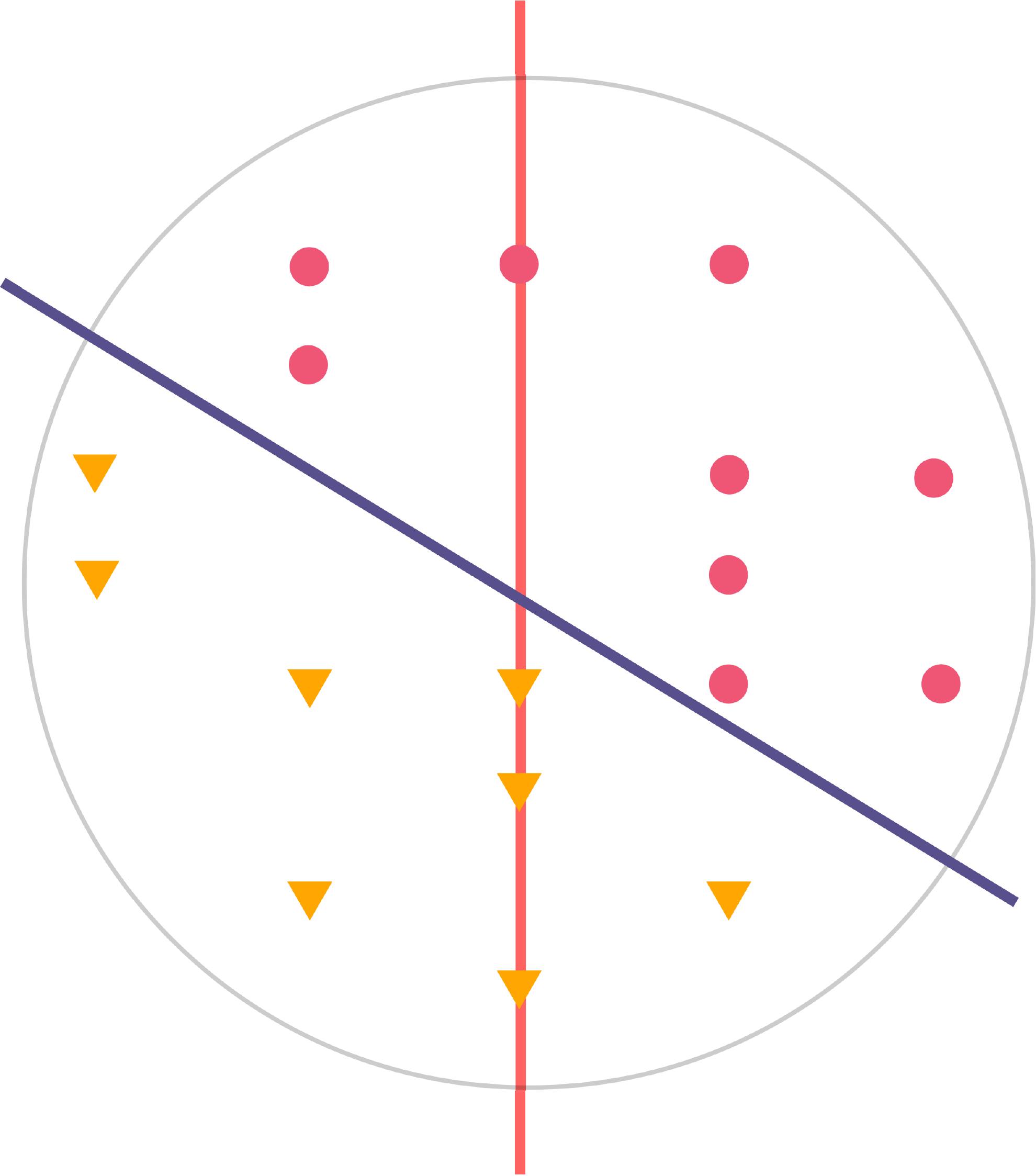}
\caption{$\good$ (triangles) and $\bad$ (circles).
No horizontal or vertical classifier can separate $\train$,
but $\train$ is linearly separable (by a slanted classifier).}
\label{fig:separability}
\end{wrapfigure}

%% file: fig/game_tree.tex
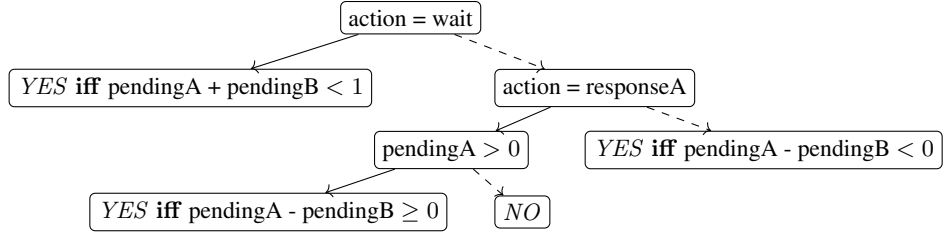
\begin{figure}
\vspace{-0mm}
\begin{center}
\begin{tikzpicture}[draw=black,rectangle,rounded corners=2pt,node distance = 4mm]

\newcommand{\ys}{6mm}

	\node (root) [draw] {action = wait};
	
	\node(d1) [below=\ys of root] {};
	\node(y) [left=4mm of d1, draw] {$\yes \mathbf{\;iff}$ pendingA + pendingB $< 1$};
	\node(n) [right=10mm of d1, draw] {action = responseA};

	\node(d2) [below=\ys of d1] {};
	\node(ny) [right=-6mm of d2, draw] {pendingA $> 0$};
	\node(nn) [right=22mm of d2, draw] {$\yes \mathbf{\;iff}$ pendingA - pendingB $< 0$};
	
	\node(d3) [below=\ys of d2] {};
	\node(nyy) [left=-6mm of d3, draw] {$\yes \mathbf{\;iff}$ pendingA - pendingB $\geq 0$};
	\node(nyn) [right=10mm of d3, draw] {$\no$};
	

	\draw [->] (root) -- node [xshift=-0.25cm, yshift=0.1cm] {} (y);
	\draw [->, dashed] (root) -- node [xshift=0.25cm, yshift=0.1cm] {} (n);
	\draw [->] (n) -- node [xshift=0.25cm, yshift=0.1cm] {} (ny);
	\draw [->, dashed] (n) -- node [xshift=-0.25cm, yshift=0.1cm] {} (nn);
	\draw [->] (ny) -- node [xshift=0.25cm, yshift=0.1cm] {} (nyy);
	\draw [->, dashed] (ny) -- node [xshift=-0.25cm, yshift=0.1cm] {} (nyn);

\end{tikzpicture}
	\caption{A decision tree for the system's controller.}
	\label{fig:game_tree}
\end{center}
\vspace{-7mm}
\end{figure}

%% file: algo/dt_linc.tex
{
\newcommand{\linclas}{c_{\vec{w},b}}
\newcommand{\sat}{\mathit{sat}}
\newcommand{\uns}{\mathit{unsat}}
\begin{algorithm}
	\caption{Learning algorithm for decision trees with linear classifiers\label{alg:dt_linc}}
	\begin{algorithmic}[1]
		\Statex \textbf{Input:} $\train \subseteq \Nats^{|\Var|}$ partitioned into subsets $\good$ and $\bad$.
		\Statex \textbf{Output:} A decision tree $\mathcal{T}$ such that $\mathcal{L}(\mathcal{T})\cap\train=\good$.
		\Statex /* train $\mathcal{T}$ on positive set $\good$ and negative set $\bad$ */
		\State $\mathcal{T} \gets (T= \set{\mathit{root}}, \rho = \emptyset, \theta = \emptyset)$
		\State $\mathsf{q} \gets \set{(\mathit{root}, \train)}$\label{line:dt_linc_qinit}
		\While{$\mathsf{q}$ nonempty}\label{line:dt_linc_loop}
			\State $(\ell, \da_\ell) \gets \mathit{pop}_{\mathsf{q}}$\label{line:dt_linc_pop}
			\If{$\da_\ell \subseteq \good$ or $\da_\ell \subseteq \bad$}\label{line:dt_linc_if_pure}
				\State $\theta(\ell) \gets \maxc(\da_\ell)$\label{line:dt_linc_pure}
			\ElsIf{$\da_\ell$ is linearly separable by a classifier $\linclas$}\label{line:dt_linc_if_sep}
				\State $\theta(\ell) \gets \linclas$\label{line:dt_linc_classifier}
			\Else\label{line:dt_linc_else}
				\State $\rho(\ell) \gets$ predicate selected by a split procedure $\mathit{Split}(\da_\ell)$\label{line:dt_linc_pred}
				\State create children $\ell_\sat$ and $\ell_\uns$ of $\ell$\label{line:dt_linc_children}
				\State $\mathit{push}_\mathsf{q}((\ell_\sat, \da_\ell[\rho(\ell)]))$, \,$\mathit{push}_\mathsf{q}((\ell_\uns, \da_\ell[\neg\rho(\ell)]))$\label{line:dt_linc_push}
			\EndIf
		\EndWhile
		\State\Return $\mathcal{T}$
	\end{algorithmic}
\end{algorithm}
}

%% file: 4_example.tex

\begin{figure}
	\begin{subfigure}[b]{0.45\textwidth}
		\centering
		\includegraphics[scale=0.5]{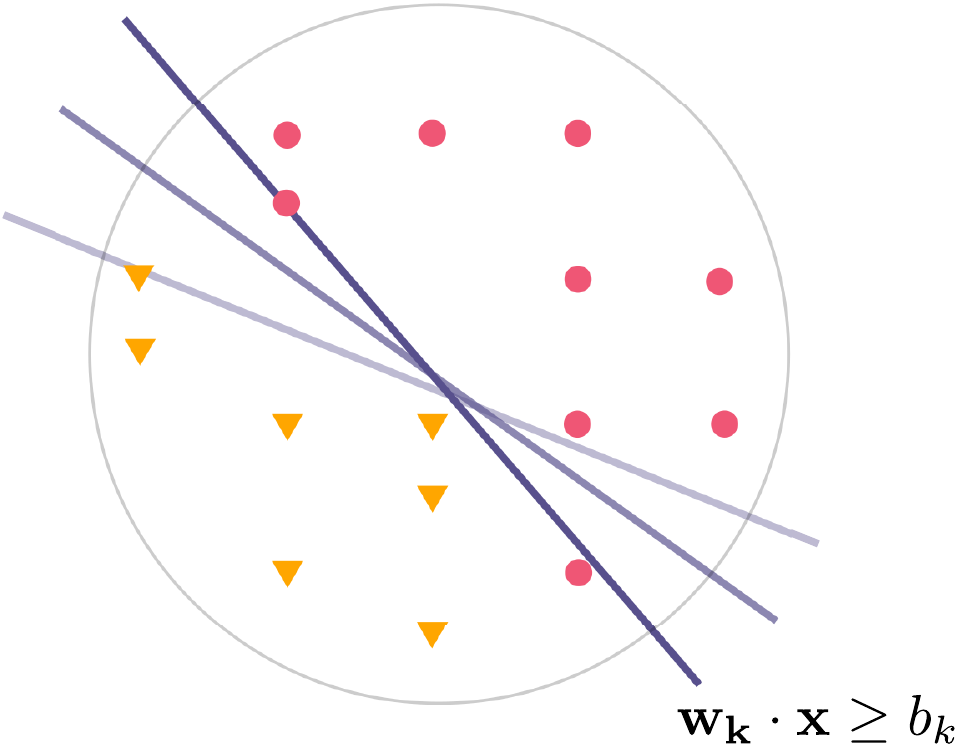}
		\caption{\footnotesize\textbf{Step 1}: We start with the entire input training set $\train$. In Line~\ref{line:dt_linc_if_pure}, we check whether $\train$ is homogeneous, and it is not. Then in Line~\ref{line:dt_linc_if_sep}, we check whether $\train$ is linearly separable, and it is not. Thus the root node becomes an inner node.}
		\vspace{2em}
	\end{subfigure}
	\hfill
	\begin{subfigure}[b]{0.45\textwidth}
		\centering
		\includegraphics[scale=0.5]{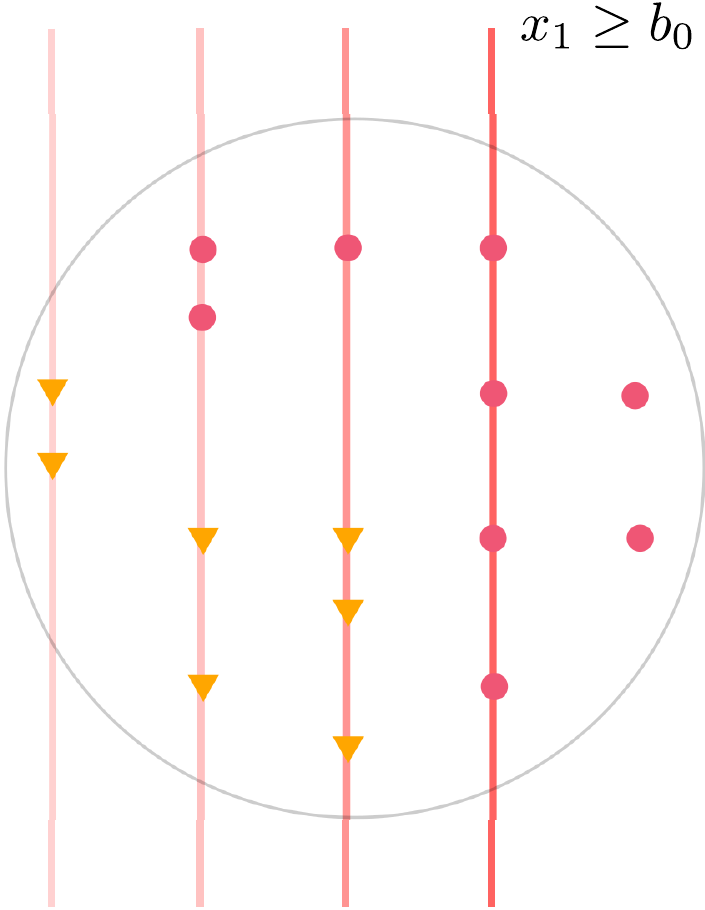}
		\caption{\footnotesize\textbf{Step 2}: An external procedure proposes a predicate for the root node (Line~\ref{line:dt_linc_pred}), and $\train$ is partitioned based on the predicate. Two children with the corresponding partitions are created (Line~\ref{line:dt_linc_push}), they will need to be processed.}
		\vspace{2em}
	\end{subfigure}
	\begin{subfigure}[b]{0.45\textwidth}
		\centering
		\includegraphics[scale=0.5]{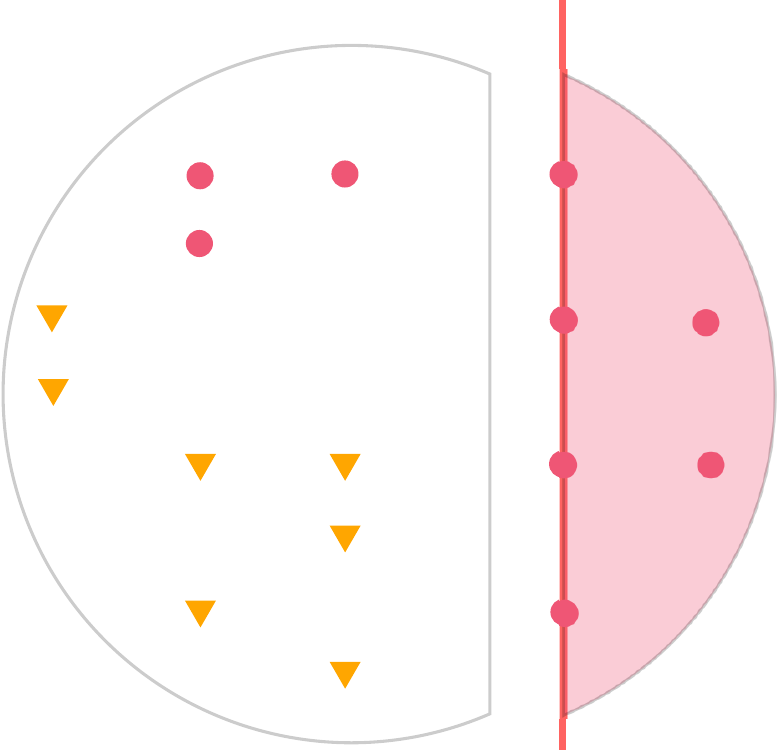}
		\caption{\footnotesize\textbf{Step 3}: Further iterations of the main loop (Line~\ref{line:dt_linc_loop}) process the two created children. One child contains a homogeneous dataset. Hence in the iteration when it is processed, in Line~\ref{line:dt_linc_pure} it becomes a pure leaf node.}
	\end{subfigure}
	\hfill
	\begin{subfigure}[b]{0.45\textwidth}
		\centering
		\includegraphics[scale=0.5]{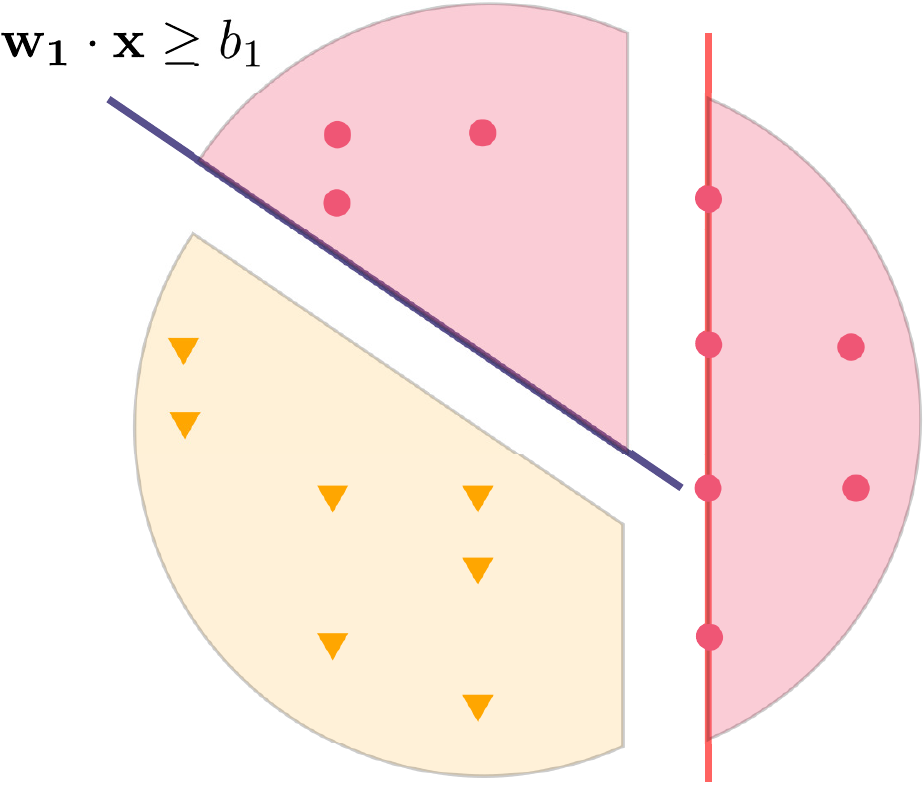}
		\caption{\footnotesize\textbf{Step 4}: The second child has a dataset which is not homogeneous, but it is linearly separable. Thus in its iteration, in Line~\ref{line:dt_linc_classifier} it becomes a classifier leaf node. No more nodes are left to be processed and so the algorithm concludes.}
	\end{subfigure}
	\hfill
	\caption{Running Algorithm \ref{alg:dt_linc} on a sample dataset consisting of circle ($Bad$) and triangle ($Good$) points. The Decision tree thus obtained is depicted in Fig. \ref{fig:running-example-dt}.}\label{fig:running-example-flow}
\end{figure}

\begin{figure}
	\centering
	\includegraphics[scale=0.6]{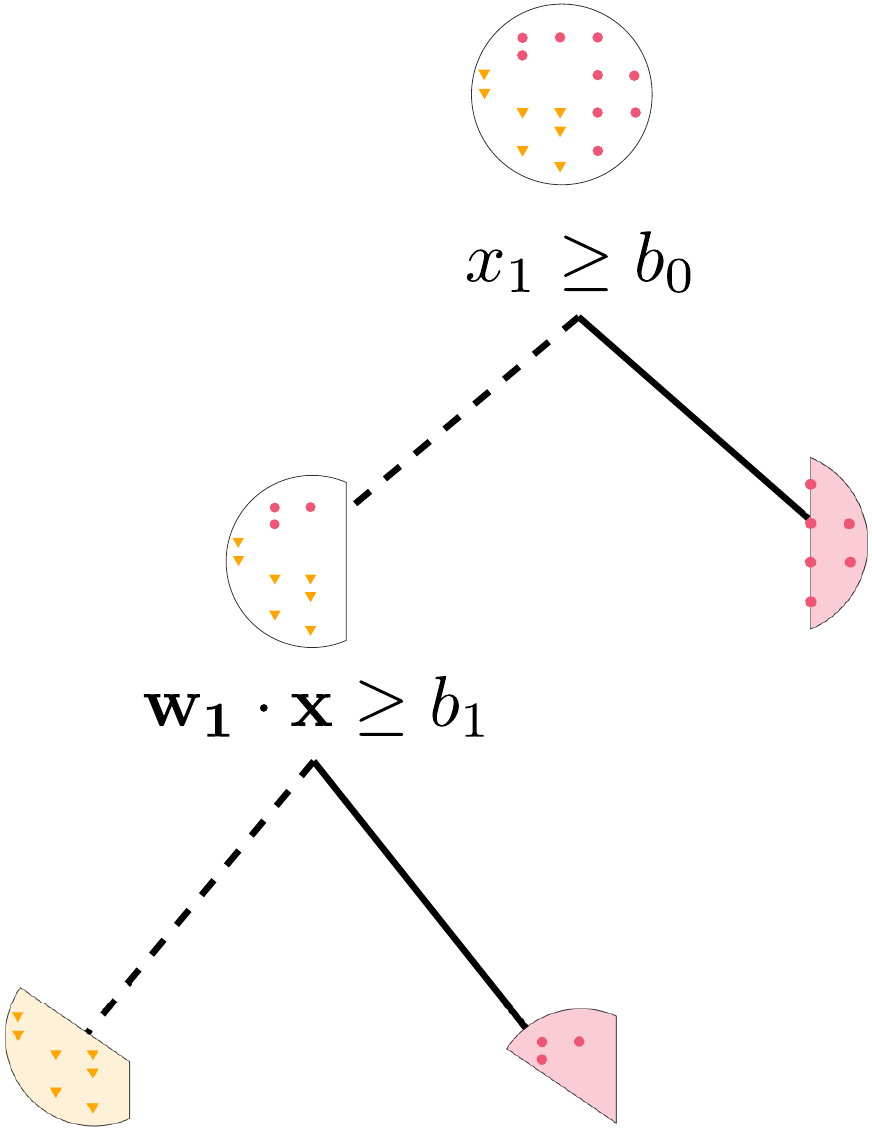}
	\caption{The Decision tree resulting from the above steps.}\label{fig:running-example-dt}
\end{figure}

\smallskip\noindent{\bf Construction of decision trees with linear classifiers.}
We present a simple running example that illustrates the key points of~\cref{alg:dt_linc}.
\cref{fig:running-example-flow} captures the flow of construction and
\cref{fig:running-example-dt} presents the output decision tree.

%% file: 4_proof.tex

\smallskip\noindent{\bf Correctness.}
We now prove the correctness of \cref{alg:dt_linc}. In other words, we show
that given a strategy in the form of a training set, \cref{alg:dt_linc} can be used to
provably represent the training set (i.e., the strategy) without errors.

\begin{theorem}
Let $G$ be a stochastic graph game, and let $\strat \colon S_i \to A_i$ be a memoryless
strategy for player $i \in \{1, 2\}$ that defines a training set $\train$ partitioned into $\good$ and $\bad$.
Consider an arbitrary split procedure that considers
only predicates from $\preds$
which produce nonempty sat- and unsat-partitions.
Given $\train$ as input, Algorithm~\ref{alg:dt_linc} using the split procedure outputs
a decision tree $\mathcal{T} = (T, \rho, \theta)$ such that $\mathcal{L}(\mathcal{T}) \cap \train = \good$,
which means that for all $s\in S_i$ we have that $\tuple{s,a}\in \mathcal{L}(\mathcal{T})$ iff $\strat(s)=a$.
Thus $\mathcal{T}$ represents the strategy $\strat$.
\end{theorem}

\newcommand{\linclas}{c_{\vec{w},b}}

\begin{proof}
We consider stochastic graph games with variables $\Var$ over a finite domain $X \subseteq \Nats$,
thus $\train \subseteq X^{|\Var|}$.
Recall that given a decision tree $\mathcal{T}=(T, \rho, \theta)$ constructed by Algorithm~\ref{alg:dt_linc},
$\rho$ assigns to every inner node a predicate from $\preds$,
and $\theta$ assigns to every leaf either $\yes$, or $\no$, or a linear classifier $\linclas$ that classifies
elements from $\Reals^{|\Var|}$ into $\yes$ resp. $\no$.

\smallskip\noindent{\em Partial correctness.}
Consider Algorithm~\ref{alg:dt_linc} with input $\train$, and let $\mathcal{T} = (T, \rho, \theta)$ be the output decision tree.
Consider an arbitrary $\tuple{s,a} \in S_i\times A_i$, note that it belongs to $\train$. Consider the leaf $\ell$ corresponding
to $\tuple{s,a}$ in $\mathcal{T}$. There is a unique path for $\tuple{s,a}$ down the tree $\mathcal{T}$ from its root,
induced by the predicates in the inner nodes given by $\rho$. Thus $\ell$ is well-defined.
At some point during the algorithm, $\ell$ was popped from the queue $\mathsf{q}$ in Line~\ref{line:dt_linc_pop}, together with
a dataset $\da_\ell$, and note that $\tuple{s,a} \in \da_\ell$. Since $\ell$ is a leaf, there are three cases
to consider:
\begin{compactenum}
\item $\theta(\ell) = \yes$. Then $\da_\ell \subseteq \mathcal{L}(\mathcal{T})$, which implies $\tuple{s,a} \in \mathcal{L}(\mathcal{T})$.
The assignment happened in Line~\ref{line:dt_linc_pure}, so (i) the condition in Line~\ref{line:dt_linc_if_pure}
was satisfied, and (ii) $\maxc(\da_\ell) = \yes$. Thus $\da_\ell \subseteq \good$, which implies $\tuple{s,a} \in \good$.
By the definition of $\good$, we have $\strat(s) = a$.
\item $\theta(\ell) = \no$. Then $\da_\ell \cap \mathcal{L}(\mathcal{T}) = \emptyset$, which implies $\tuple{s,a} \not\in \mathcal{L}(\mathcal{T})$.
The assignment happened in Line~\ref{line:dt_linc_pure}, so (i) the condition in Line~\ref{line:dt_linc_if_pure}
was satisfied, and (ii) $\maxc(\da_\ell) = \no$. Thus $\da_\ell \subseteq \bad$, which implies $\tuple{s,a} \in \bad$.
By the definition of $\bad$, we have $\strat(s) \not= a$.
\item $\theta(\ell) = \linclas$. This assignment happened in Line~\ref{line:dt_linc_classifier}. Thus the condition in
Line~\ref{line:dt_linc_if_sep} was satisfied, and hence $\linclas$ linearly separates $\da_\ell$. As $\tuple{s,a} \in \da_\ell$,
we have that $\linclas(\tuple{s,a}) = \yes$ iff $\tuple{s,a} \in \good$. This gives that $\tuple{s,a} \in \mathcal{L}(\mathcal{T})$
iff $\strat(s) = a$.
\end{compactenum}
The desired result follows.

\smallskip\noindent{\em Total correctness.}
Algorithm~\ref{alg:dt_linc} uses a split procedure that considers
only predicates from $\preds$
which produce nonempty sat- and unsat-partitions.
Thus the algorithm maintains the following invariant for every path $\bar{p}$ in $\mathcal{T}$ starting from the root:
For each predicate $\pred \in \preds$,
there is at most one inner node $\bar{n}$ in the path $\bar{p}$ such that $\rho(\bar{n}) = \pred$.
This invariant is indeed maintained, since any predicate considered the second time in a path inadvertedly produces
an empty data partition, and such predicates are not considered by the split procedure that
selects predicates for $\rho$ (in Line~\ref{line:dt_linc_pred} of Algorithm~\ref{alg:dt_linc}).

From the above we have that the length of any path in $\mathcal{T}$ starting from the root is at most
$|\preds| \leq 2 \cdot |\Var| \cdot |X|$, i.e., twice the number of variables times the size of the variable domain.
We prove that the number of iterations of the loop in Line~\ref{line:dt_linc_loop} is finite.
The branch from Line~\ref{line:dt_linc_else} happens finitely many times, since it adds two vertices (in Line~\ref{line:dt_linc_children})
to the decision tree $\mathcal{T}$ and we have the bound on the path lengths in $\mathcal{T}$.
Since only the branch from Line~\ref{line:dt_linc_else} pushes elements into the queue $\mathsf{q}$,
and each iteration of the loop pops an element from $\mathsf{q}$ in Line~\ref{line:dt_linc_pop},
the number of loop iterations (Line~\ref{line:dt_linc_loop}) is indeed finite.
This proves termination, which together with partial correctness proves total correctness.\qed

\end{proof}

%% file: fig/truefalse_posneg.tex
\begin{wrapfigure}[7]{r}{0.45\textwidth}
\vspace{-0mm}
\centering
\small
\begin{tikzpicture}[thick, >=latex,
pre/.style={<-,shorten >= 1pt, shorten <=1pt, thick},
post/.style={->,shorten >= 1pt, shorten <=1pt,  thick},
und/.style={very thick, draw=gray},
node1/.style={circle, minimum size=4mm, draw=black!100, line width=1pt, inner sep=0},
node2/.style={circle, minimum size=5mm, draw=black!100, fill=white!100, very thick, inner sep=0},
nodeempty/.style={},
virt/.style={circle,draw=black!50,fill=black!20, opacity=0}]

\newcommand{\myblue}{blue!85!black}
\newcommand{\myred}{red!85!black}
\newcommand{\mygreen}{green!70!black}
\newcommand{\xstep}{1.2}
\newcommand{\ystep}{0.6}


\node[nodeempty] (c) at (0.3*\xstep, 0.1*\ystep){$c$};
\node[nodeempty] (y) at (1*\xstep, 0){$\yes$};
\node[nodeempty] (n) at (2*\xstep, 0){$\no$};

\node[nodeempty] (x) at (-0.2*\xstep, -0.2*\ystep){$\vec{x}$};
\node[nodeempty] (g) at (0, -1*\ystep){$\good$};
\node[nodeempty] (b) at (0, -2*\ystep){$\bad$};

\draw[-, very thick] (-0.5*\xstep, 0.5*\ystep) to (2.5*\xstep, 0.5*\ystep);
\draw[-, very thick] (-0.5*\xstep, -0.5*\ystep) to (2.5*\xstep, -0.5*\ystep);
\draw[-, very thick] (-0.5*\xstep, -1.5*\ystep) to (2.5*\xstep, -1.5*\ystep);
\draw[-, very thick] (-0.5*\xstep, -2.5*\ystep) to (2.5*\xstep, -2.5*\ystep);

\draw[-, very thick] (-0.5*\xstep, 0.5*\ystep) to (-0.5*\xstep, -2.5*\ystep);
\draw[-, very thick] (0.5*\xstep, 0.5*\ystep) to (0.5*\xstep, -2.5*\ystep);
\draw[-, very thick] (1.5*\xstep, 0.5*\ystep) to (1.5*\xstep, -2.5*\ystep);
\draw[-, very thick] (2.5*\xstep, 0.5*\ystep) to (2.5*\xstep, -2.5*\ystep);

\draw[-, thick] (-0.5*\xstep, 0.5*\ystep) to (0.5*\xstep, -0.5*\ystep);

\node[nodeempty] (tp) at (1*\xstep, -1*\ystep){$\tp$};
\node[nodeempty] (fn) at (2*\xstep, -1*\ystep){$\fn$};
\node[nodeempty] (fp) at (1*\xstep, -2*\ystep){$\fp$};
\node[nodeempty] (tn) at (2*\xstep, -2*\ystep){$\tn$};




\end{tikzpicture}
\caption{True/False Positive/Negative.}
\label{fig:truefalse_posneg}
\end{wrapfigure}

%% file: fig/roc.tex
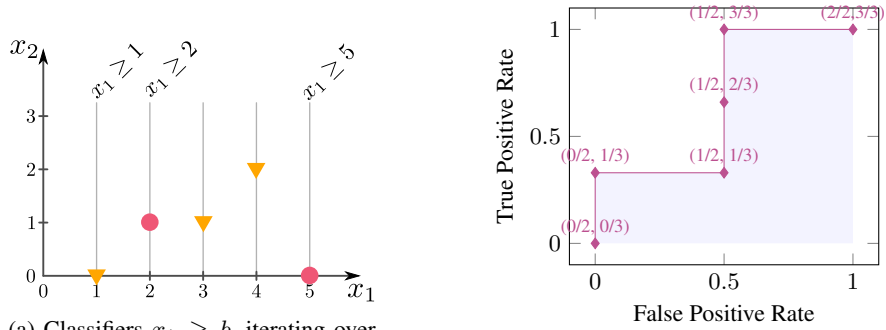
\begin{figure}
	\centering
	\begin{subfigure}[b]{0.4\textwidth}
		\centering
		\includegraphics[scale=0.7]{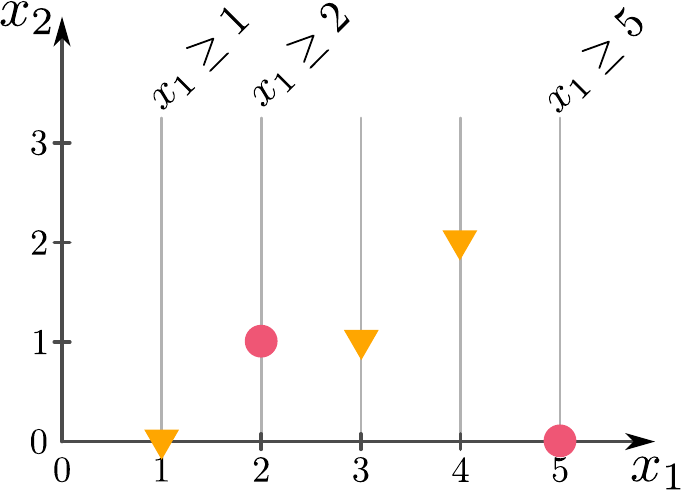}
		\caption{Classifiers $x_1 \geq b$, iterating over the bias $b$ from $5$ down to $1$.}
	\end{subfigure}
	\hfill
	\begin{subfigure}[b]{0.5\textwidth}
		\centering
		\begin{tikzpicture}
			\begin{axis}[
			xlabel near ticks,
			ylabel near ticks,
			xlabel={False Positive Rate},
			ylabel={True Positive Rate},
			scale=0.6,
			]
			\definecolor{mycolor}{RGB}{188,80,144}
			\addplot [name path=roc,mycolor,mark=diamond*,point meta=explicit symbolic, nodes near coords] coordinates { (0.0, 0.0)    [\scriptsize (0/2, 0/3)]
				(0.0, 0.33)	  [\scriptsize (0/2, 1/3)]
				(0.5, 0.33)   [\scriptsize (1/2, 1/3)]
				(0.5, 0.66)    [\scriptsize (1/2, 2/3)]
				(0.5, 1.0)    [\scriptsize (1/2, 3/3)]
				(1.0, 1.0)    [\scriptsize (2/2,3/3)]};
			\path[name path=axis] (axis cs:0,0) -- (axis cs:1,0);
			\addplot [
				thick,
				color=mycolor,
				fill=blue,
				fill opacity=0.05
			]
			fill between [
				of=roc and axis,
				soft clip={domain=0:1},
			];
			\end{axis}
		\end{tikzpicture}
		\caption{ROC curve and the shaded $\auc$.}
	\end{subfigure}
	\vspace{-2mm}
	\caption{Area under the curve for $\vec{w}=(1,0)$ w.r.t. $\good$ (triangles) and $\bad$ (circles).}
	\label{fig:roc}
\end{figure}

%% file: algo/split_auc.tex
{
\newcommand{\wg}{\vec{w}}
\newcommand{\sat}{\mathit{sat}}
\newcommand{\uns}{\mathit{unsat}}
\newcommand{\areas}{\mathsf{areas}}
\newcommand{\lls}{\mathit{LinearLeastSquares}}
\begin{algorithm}
	\caption{Split procedure -- area under the curve ($\auc$)\label{alg:split_auc}}
	\begin{algorithmic}[1]
		\Statex \textbf{Input:} $\da \subseteq \Nats^{|\Var|}$ partitioned into subsets $\da_G$ and $\da_B$.
		\Statex \textbf{Output:} A predicate $\pred$ maximizing area under the $\sat$ and $\uns$ ROC curves.
		\State $\areas \gets \emptyset$
		\For{$\pred \in \preds$}\label{line:split_auc_loop}
			\State $\wg_\sat \gets \lls(\da[\pred])$\label{line:split_auc_sat}
			\State $\wg_\uns \gets \lls(\da[\neg\pred])$\label{line:split_auc_unsat}
			\State $\areas(\pred) \gets \auc(\wg_\sat, \da[\pred]) + \auc(\wg_\uns, \da[\neg\pred])$\label{line:split_auc_score}
		\EndFor
		\State\Return $\argmax_{\pred} \set{\areas(\pred)}$\label{line:split_auc_ret}
	\end{algorithmic}
\end{algorithm}
}

%% file: 5_experiments.tex
\section{Experiments}\label{sec:exper}

Throughout our experiments, we consider the following construction algorithms:
\begin{compactitem}
\item Basic decision trees (\cref{alg:dt_basic}~with~\cref{alg:split_ig}), as considered in~\cite{DBLP:conf/tacas/BrazdilCKT18}. \hfill $(\star)$
\item Decision trees with linear classifiers (\cref{alg:dt_linc}) and entropy-based splitting procedure (\cref{alg:split_ig}). \hfill $(\dagger)$
\item Decision trees with linear classifiers (\cref{alg:dt_linc}) and auc-based splitting procedure (\cref{alg:split_auc}). \hfill $(\ddagger)$
\end{compactitem}

For the experimental evaluation of the construction algorithms, we consider multiple sources of problems that
arise naturally in reactive synthesis, and reduce to stochastic graph games with Integer variables.
These variables provide semantical information about the states (resp., actions) they identify,
so a strategy-representation method utilizing predicates over the variables produces naturally interpretable output.
Moreover, there is an inherent internal structure in the states and their valuations, which machine-learning algorithms
can exploit to produce more succinct representation of strategies.

Given a game and an objective, we use an explicit solver to obtain an almost-sure winning strategy.
Then we consider the strategy as a list of played ($\good$) and non-played ($\bad$) actions for each state, which
can be used directly as an input training set ($\train$). We evaluate the construction algorithms based on succinctness
of representation, which we express as the number of non-pure nodes (i.e., nodes with either a predicate or a linear classifier).
Further experimental details are presented in~\cref{app:exp}. 

\vspace{-2mm}
\subsection{Graph games and winning strategies}

We consider two sources of problems reducible to strategy representation in graph games,
namely, AIGER safety synthesis~\cite{AIGER} and LTL synthesis~\cite{Pnueli77}.

\vspace{-3mm}
\subsubsection{AIGER -- Scheduling of Washing Cycles.}
The goal of this problem is to design a centralized controller for a system of washing tanks running in
parallel. The system is parametrized by the number of tanks, the time limit to fill a tank with water
after a request, the delay after which the tank has to be emptied again, and a number of tanks per one
shared water pipe. The controller has to ensure that all requests are satisfied within the specified time limit.

The problem has been introduced in the second year of SYNTCOMP~\cite{DBLP:journals/corr/JacobsBBKPRRSST16},
the most important and well-known synthesis competition. The problem is implicitly described in the form
of AIGER safety specification~\cite{AIGER}, which uses circuits with input, output, and latch Boolean variables.
This reduces directly to graph games with $\set{0,1}$-valued Integer variables and safety objectives.
The state-variables represent for each tank whether it is currently filled, and the current deadline for filling (resp., emptying).
The action-variables capture environment requests to fill water tanks,
and the controller commands to fill (resp., empty) water tanks.
We consider 364 datasets, where the sizes of $\train$ range from 640 to 1024000, and
the sizes of $\Var$ range from 16 to 62.

\begin{figure}
\begin{minipage}{.51\textwidth}
	\begin{center}
	\includegraphics[scale=0.24]{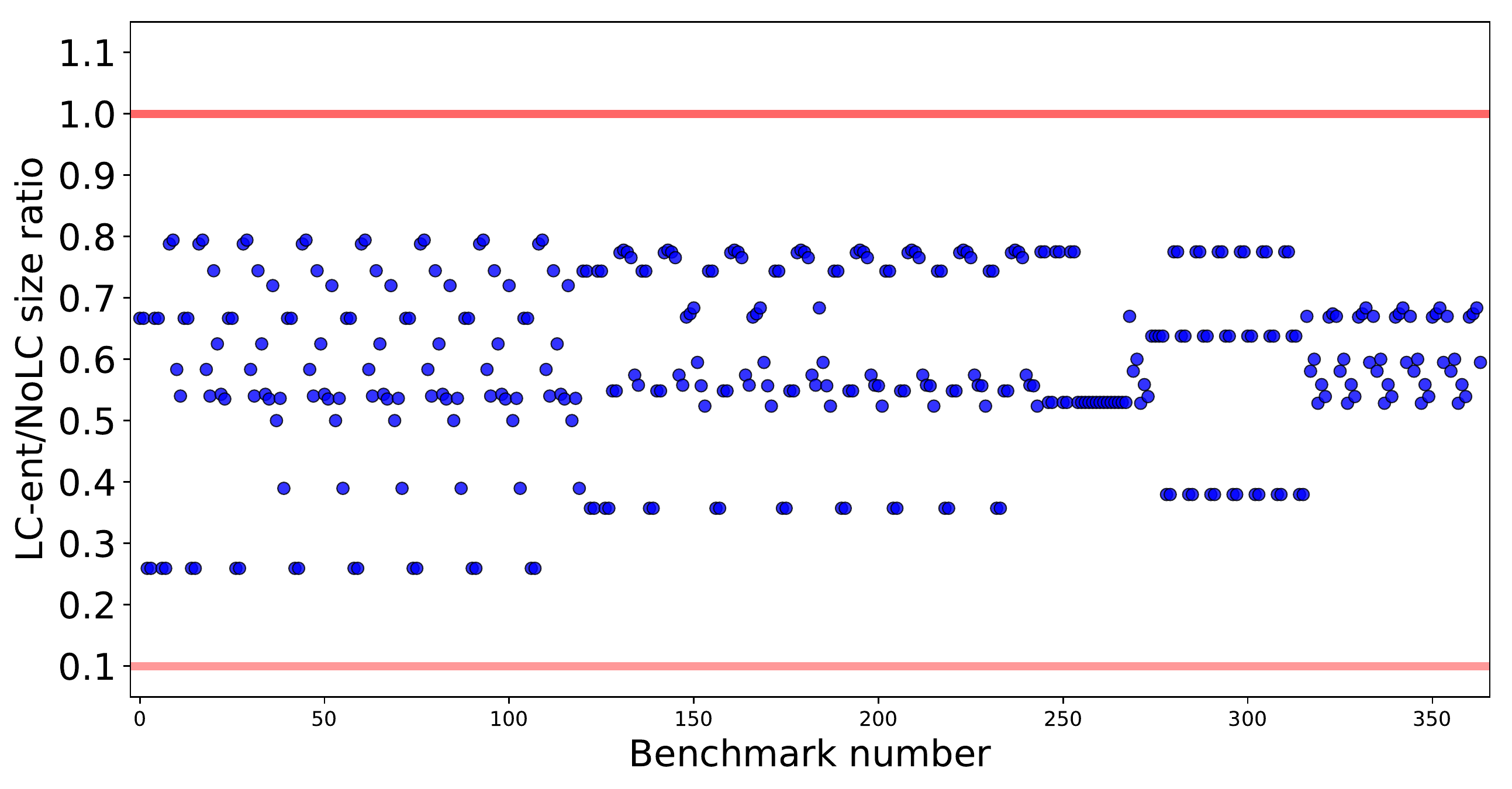}
	\end{center}
\end{minipage}
\begin{minipage}{.5\textwidth}
	\begin{center}
	\includegraphics[scale=0.24]{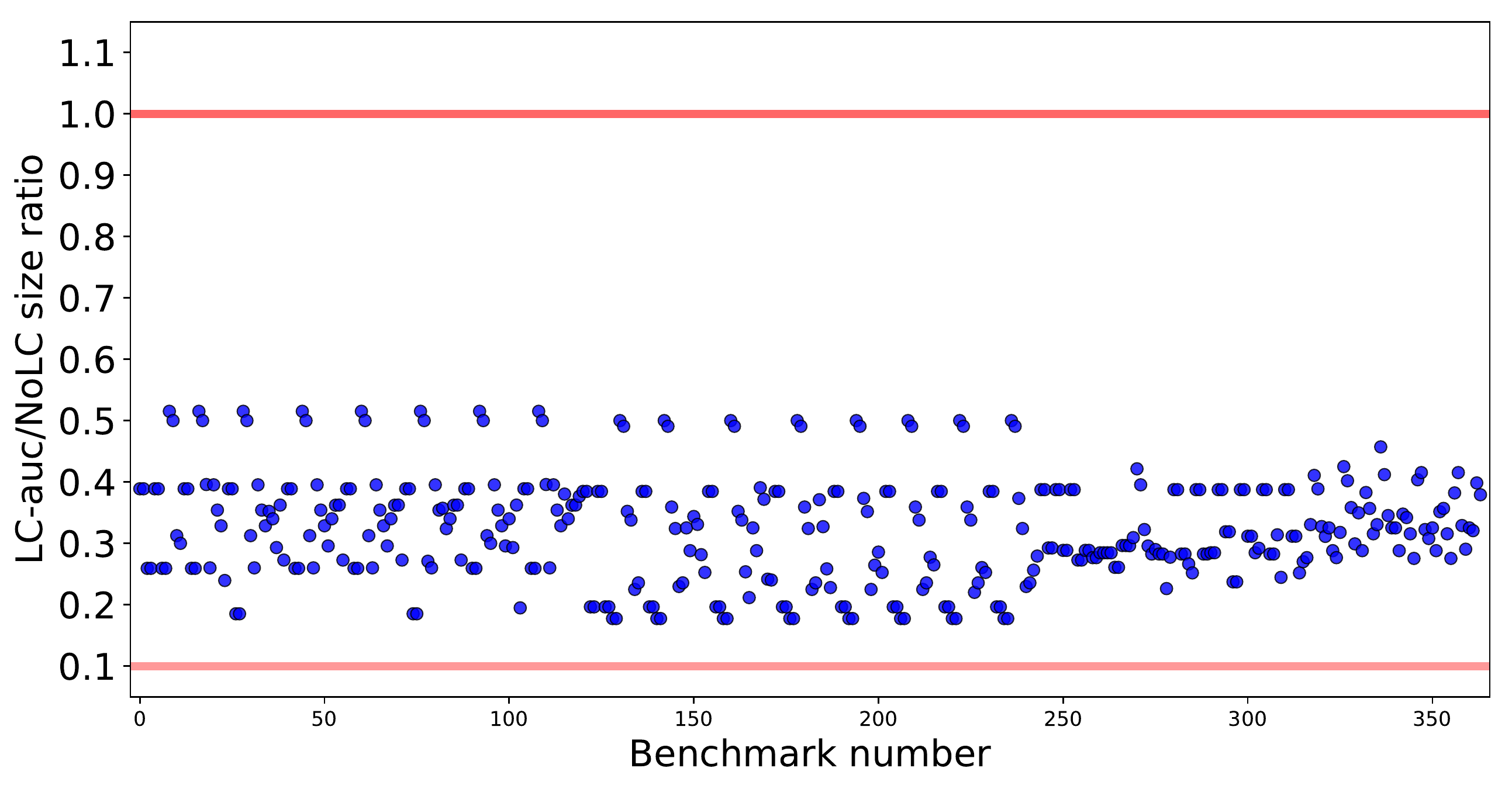}
	\end{center}
\end{minipage}
\caption{Scheduling of Washing Cycles.}
\label{fig:wash}
\end{figure}
\vspace{-1mm}
We illustrate the results in \cref{fig:wash}. Both subfigures plot the ratios of sizes for two considered
algorithms. Each dot represents a dataset, the $y$-axis captures the ratios, and the two red lines represent
equality and order-of-magnitude improvement, respectively. The left figure considers the size ratios of the
basic decision-tree algorithm and the algorithm with linear classifiers and entropy-based splits ($\star/\dagger$).
The arithmetic, geometric, and harmonic means of the ratios are $59\%$, $57\%$, and $55\%$, respectively.
The right figure considers the basic algorithm and the algorithm with linear classifiers and auc-based splits ($\star/\ddagger$).
The arithmetic, geometric, and harmonic means of the ratios are $33\%$, $31\%$, and $30\%$, respectively.

\vspace{-3mm}
\subsubsection{LTL synthesis.}

In reactive synthesis, most properties considered in practice are \mbox{$\omega$-regular} objectives,
which can be specified as linear-time temporal logic (LTL) formulae over input/output signals~\cite{Pnueli77}.
Given an LTL formula and input/output signal partitioning, the controller synthesis for this specification
is reducible to solving a graph game with parity objective.

In our experiments, we consider LTL formulae randomly generated using the tool SPOT~\cite{spot}.
Then, we use the tool Rabinizer~\cite{rabinizer} to translate the formulae into deterministic parity automata.
Crucially, the states of these automata contain semantic information retained by Rabinizer during
the translation. We consider an encoding of the semantic information
(given as sets of LTL formulae and permutations) into binary vectors. The encoding aims to capture
the inherent structure within automaton states, which can later be exploited during strategy representation.
Finally, for each parity automaton we consider various input/output partitionings of signals,
and thus we obtain parity graph games with $\set{0,1}$-valued Integer variables.
The whole pipeline is described in detail in~\cite{DBLP:conf/tacas/BrazdilCKT18}.

We consider graph games with liveness (parity-2) and strong fairness (parity-3) objectives.
In total we consider 917 datasets, with sizes of $\train$ ranging from 48 to 8608, and sizes of $\Var$ ranging from 38 to 128.

\begin{figure}
\begin{minipage}{.51\textwidth}
	\begin{center}
	\includegraphics[scale=0.24]{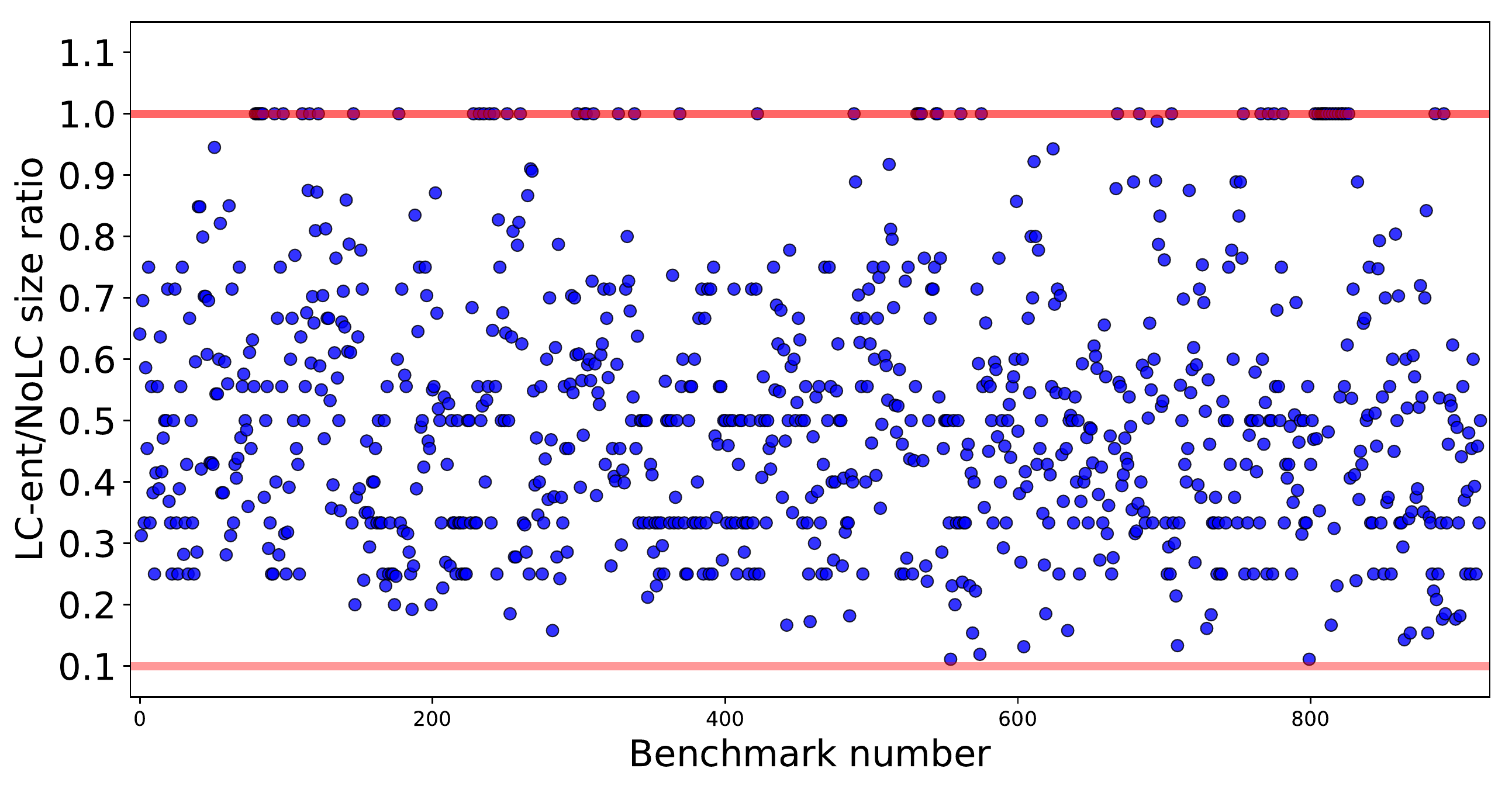}
	\end{center}
\end{minipage}
\begin{minipage}{.5\textwidth}
	\begin{center}
	\includegraphics[scale=0.24]{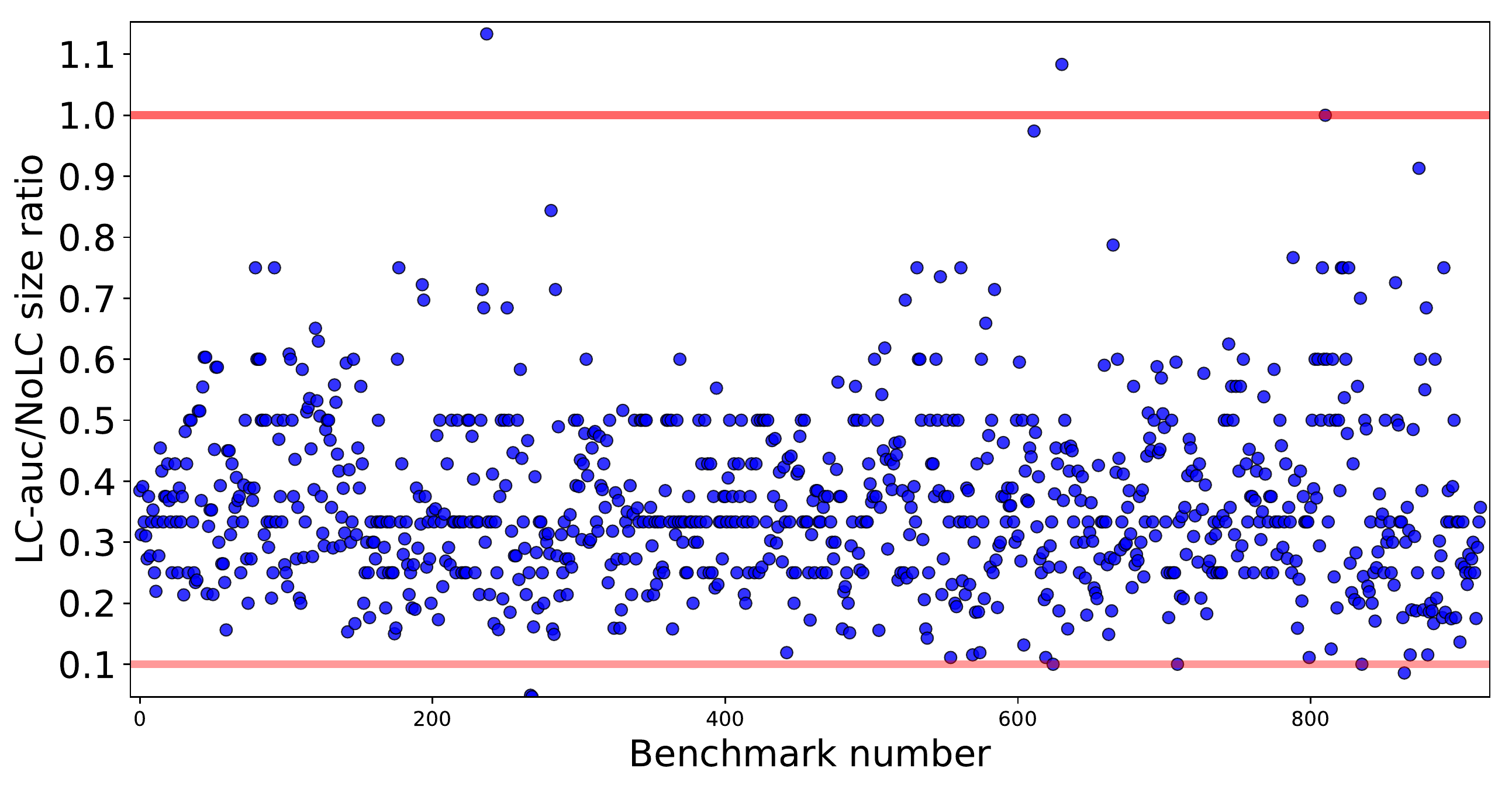}
	\end{center}
\end{minipage}
\caption{LTL synthesis.}
\label{fig:ltlgames}
\end{figure}
\vspace{-1mm}
\cref{fig:ltlgames} illustrates the results, where both subfigures plot the ratios of sizes
(captured on the $y$-axis) for two considered algorithms. The left figure considers the
basic decision-tree algorithm and the algorithm with linear classifiers and entropy-based splits ($\star/\dagger$).
The arithmetic, geometric, and harmonic means of the ratios are $51\%$, $47\%$, and $43\%$, respectively.
The right figure considers the basic decision-tree algorithm and the algorithm with linear classifiers and auc-based splits ($\star/\ddagger$).
The arithmetic, geometric, and harmonic means of the ratios are $36\%$, $34\%$, and $31\%$, respectively.

\subsection{MDPs and almost-sure winning strategies}
\subsubsection{LTL synthesis with randomized environment.}

In LTL synthesis, given a formula and an input/otput signal partitioning, there may be no controller that
satisfies the LTL specification. In such a case, it is natural to consider a different setting where the environment
is not antagonistic, but behaves randomly instead. There are LTL specifications that
are unsatisfiable, but become satisfiable when randomized environment is considered. Such special
case of LTL synthesis reduces to solving MDPs with almost-sure parity objectives~\cite{DBLP:journals/jacm/ChatterjeeHJ015}.
Note that in this setting, the precise probabilities of environment actions are immaterial, as they have no effect
on the existence of a controller ensuring an objective almost-surely (i.e., with probability $1$).

We consider 414 instances of LTL synthesis reducible to graph games with \emph{co-B\"{u}chi} (i.e., parity-$2$)
objective, where the LTL specification is unsatisfiable, but becomes satisfiable with randomized environment
(which reduces to MDPs with almost-sure co-B\"{u}chi objective).
The examples have been obtained by the same pipeline as the one described in the previous subsection.
In the examples, the sizes of $\train$ range from 80 to 26592, and the sizes of $\Var$ range from 38 to 74.

\begin{figure}
\begin{minipage}{.51\textwidth}
	\begin{center}
	\includegraphics[scale=0.24]{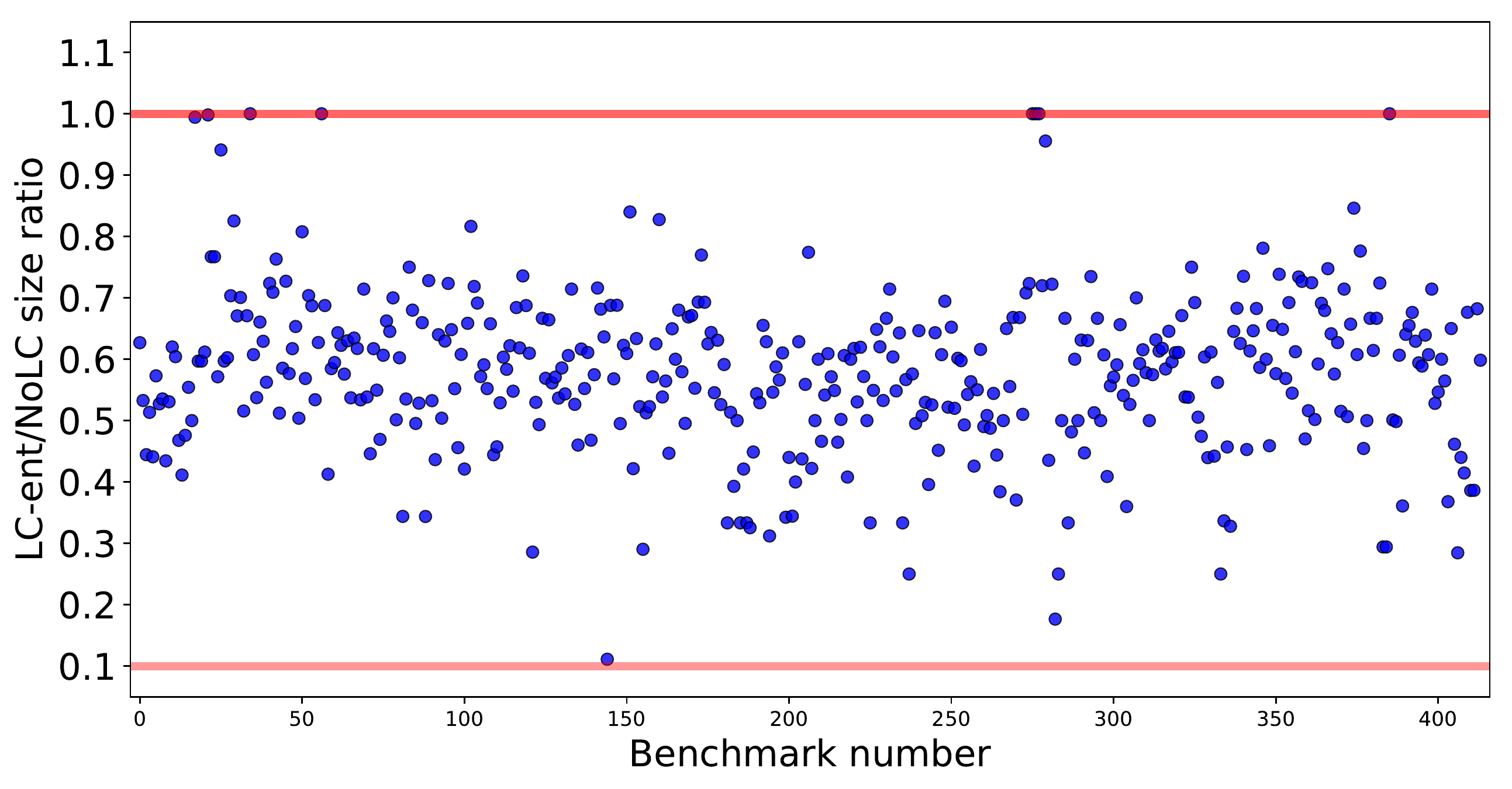}
	\end{center}
\end{minipage}
\begin{minipage}{.5\textwidth}
	\begin{center}
	\includegraphics[scale=0.24]{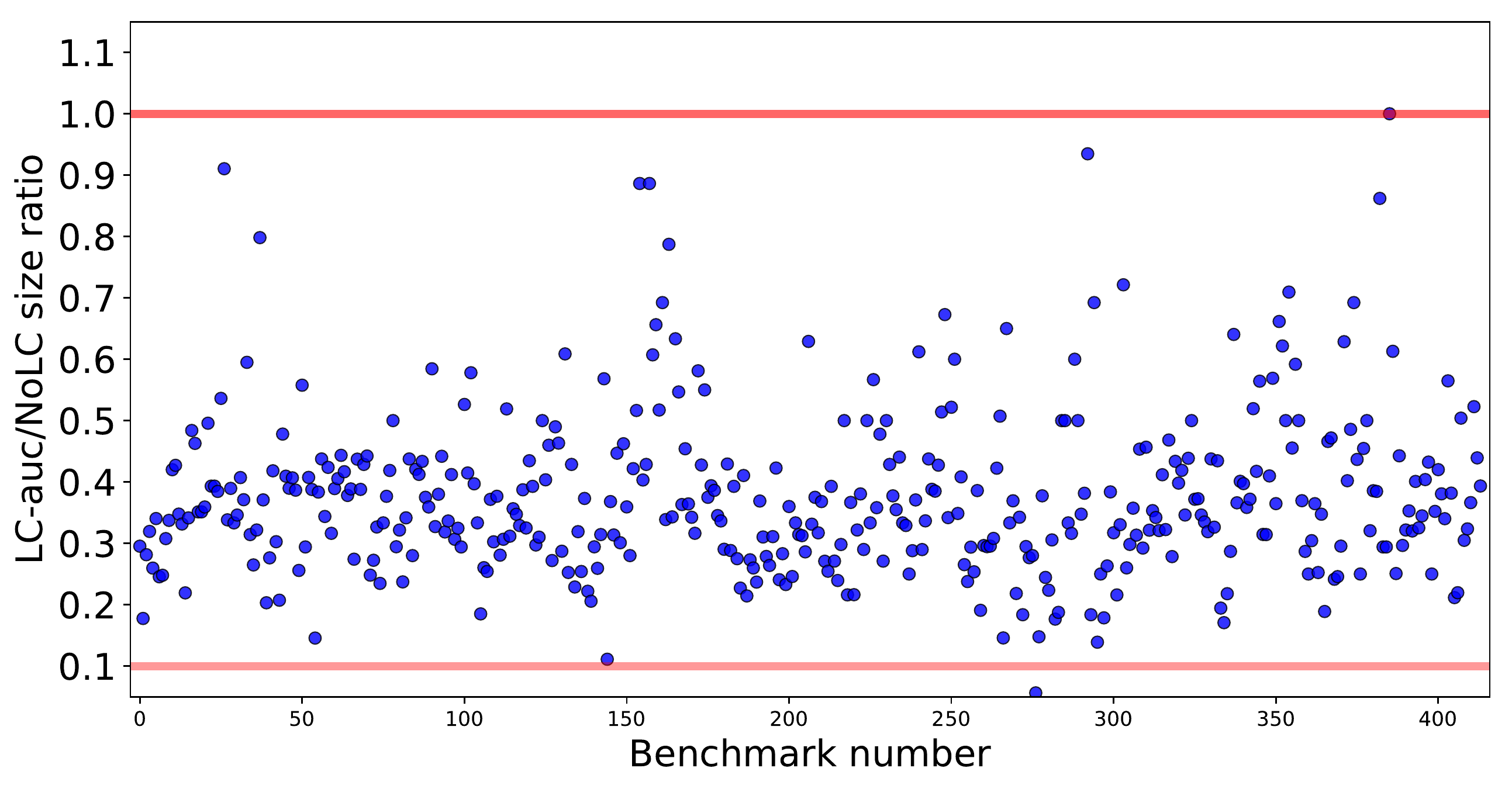}
	\end{center}
\end{minipage}
\caption{LTL synthesis with randomized environment.}
\label{fig:ltlmdps}
\end{figure}
The experimental results are summarized in \cref{fig:ltlmdps}. The two subfigures plot the ratios of sizes
(captured on the $y$-axis) for two considered algorithms. The left figure considers the
basic decision-tree algorithm and the algorithm with linear classifiers and entropy-based splits ($\star/\dagger$).
The arithmetic, geometric, and harmonic means of the ratios are $58\%$, $56\%$, and $54\%$, respectively.
The right figure considers the basic decision-tree algorithm and the algorithm with linear classifiers and auc-based splits ($\star/\ddagger$).
The arithmetic, geometric, and harmonic means of the ratios are $38\%$, $36\%$, and $34\%$, respectively.


\subsubsection{PRISM model checking.}
We consider model checking of probabilistic systems in the model checker PRISM~\cite{PRISM}.
Given an implicit description of a probabilistic system in PRISM, and a reachability/safety LTL formula as
a specification, the model checking problem of the model and the specification reduces to
construction of an almost-sure winning strategy in an MDP with nonnegative Integer variables.
The state-variables correspond to the variables in the implicit PRISM model description,
i.e., local states of the moduli, counter values, etc. The action-variables capture the id of
the module performing an action, and the id of the action performed by the module.

\begin{table}[t]
\addtocounter{table}{-1}
\input{fig/exp/prism}
\vspace{-2mm}
\caption{PRISM model checking.}
\label{tab:prism}
\end{table}

\cref{tab:prism} presents the PRISM experimental results, where we consider
various case studies available from the PRISM benchmark suite~\cite{PRISMbenchmarks}
(e.g., communication protocols).
The columns of the table represent the considered model and specification,
the sizes of $\train$ and $\Var$, and the decision-tree sizes for the three
considered construction algorithms ($\star,\dagger,\ddagger$).

In this set of experiments, we have noticed several cases where the split heuristic based on
$\auc$ achieves significantly worse results. Namely, in csma, wlan, and zeroconf, it is
mostly outperformed by the information-gain split procedure, and sometimes it is outperformed
even by standard decision trees without linear classifiers.
This was caused by certain variables repeatedly having high $\auc$ scores (for different thresholds)
when constructing some branches of the tree, even though subsequent choices of the predicates
did little progress to linearly separate the data.
We were able to mitigate
the cases of bad predicate suggestions, e.g., by penalizing the predicates on the variables that already
appear in the path to the current node (that is about to be split), however, the inferior overall
performance in these benchmarks persists. This discovery motivates to consider various combinations of
$\auc$ and information-gain methods, e.g., using information gain as a stand-in metric, in cases
where $\auc$ yields poor scores for all considered predicates.

%% file: fig/exp/prism.tex
\begin{small}
\begin{longtable}{|l | l | l | l || r | r | r |}
\hline
\textbf{Model} & \textbf{Specification} & $\mathbf{|Train|}$ & $\mathbf{|Var|}$ & $\textbf{NoLC}$ & $\textbf{LC-ent}$ & $\textbf{LC-auc}$ \\
\hline
\hline
coin2\_K1 \hspace{2mm} & F[finished\&agree] \hspace{2mm} & 1820 & 7 & 142 & 135 & \textbf{45} \\
coin2\_K2 \hspace{2mm} & F[finished\&agree] \hspace{2mm} & 3484 & 7 & 270 & 261 & \textbf{55} \\
coin2\_K3 \hspace{2mm} & F[finished\&agree] \hspace{2mm} & 5148 & 7 & 386 & 373 & \textbf{60} \\
coin2\_K4 \hspace{2mm} & F[finished\&agree] \hspace{2mm} & 6812 & 7 & 536 & 520 & \textbf{55} \\
coin2\_K9 \hspace{2mm} & F[finished\&agree] \hspace{2mm} & 15132 & 7 & 1137 & 1123 & \textbf{68} \\
coin3\_K1 \hspace{2mm} & F[finished\&agree] \hspace{2mm} & 27854 & 9 & 772 & 713 & \textbf{298} \\
coin3\_K2 \hspace{2mm} & F[finished\&agree] \hspace{2mm} & 51566 & 9 & 1142 & 1074 & \textbf{316} \\
coin3\_K3 \hspace{2mm} & F[finished\&agree] \hspace{2mm} & 75278 & 9 & 1580 & 1500 & \textbf{378} \\
coin3\_K4 \hspace{2mm} & F[finished\&agree] \hspace{2mm} & 98990 & 9 & 2047 & 1967 & \textbf{388} \\
coin4\_K0 \hspace{2mm} & F[finished\&agree] \hspace{2mm} & 52458 & 11 & 742 & 632 & \textbf{221} \\
coin5\_K0 \hspace{2mm} & F[finished\&agree] \hspace{2mm} & 451204 & 13 & 2572 & 1626 & \textbf{566} \\
\hline
csma2\_2 \hspace{2mm} & F[succ\_min\_bo$\leq$2] \hspace{2mm} & 8590 & 13 & 70 & 52 & \textbf{32} \\
csma2\_2 \hspace{2mm} & F[max\_col$\geq$3] \hspace{2mm} & 10380 & 13 & 65 & \textbf{54} & \textbf{54} \\
csma2\_3 \hspace{2mm} & F[succ\_min\_bo$\leq$3] \hspace{2mm} & 25320 & 13 & 66 & 48 & \textbf{35} \\
csma2\_3 \hspace{2mm} & F[max\_col$\geq$4] \hspace{2mm} & 28730 & 13 & 63 & \textbf{48} & 59 \\
csma2\_4 \hspace{2mm} & F[succ\_min\_bo$\leq$4] \hspace{2mm} & 73110 & 13 & 60 & 42 & \textbf{40} \\
csma2\_4 \hspace{2mm} & F[max\_col$\geq$5] \hspace{2mm} & 79580 & 13 & 54 & \textbf{41} & 59 \\
\hline
firewire\_abst \hspace{2mm} & F[exists\_leader] \hspace{2mm} & 2535 & 4 & 12 & 10 & \textbf{8} \\ 
firewire\_impl\_01 \hspace{2mm} & F[exists\_leader] \hspace{2mm} & 22633 & 12 & 99 & 86 & \textbf{71} \\
firewire\_impl\_02 \hspace{2mm} & F[exists\_leader] \hspace{2mm} & 37180 & 12 & 101 & 85 & \textbf{81} \\
firewire\_impl\_05 \hspace{2mm} & F[exists\_leader] \hspace{2mm} & 90389 & 12 & 102 & 85 & \textbf{72} \\
\hline
leader2 \hspace{2mm} & F[elected] \hspace{2mm} & 204 & 12 & 25 & 18 & \textbf{11} \\
leader3 \hspace{2mm} & F[elected] \hspace{2mm} & 3249 & 17 & 61 & 34 & \textbf{23} \\
leader4 \hspace{2mm} & F[elected] \hspace{2mm} & 38016 & 22 & 152 & 92 & \textbf{45} \\
\hline
mer10 \hspace{2mm} & G[!err\_G] \hspace{2mm} & 499632 & 19 & 552 & 510 & \textbf{124} \\ 
mer20 \hspace{2mm} & G[!err\_G] \hspace{2mm} & 954282 & 19 & 963 & 922 & \textbf{124} \\ 
mer30 \hspace{2mm} & G[!err\_G] \hspace{2mm} & 1408932 & 19 & 1373 & 1332 & \textbf{126} \\ 
\hline
wlan0 \hspace{2mm} & F[both\_sent] \hspace{2mm} & 27380 & 14 & 244 & \textbf{198} & 232 \\
wlan1 \hspace{2mm} & F[both\_sent] \hspace{2mm} & 81940 & 14 & 272 & \textbf{200} & 286 \\
wlan2 \hspace{2mm} & F[both\_sent] \hspace{2mm} & 275140 & 14 & 288 & \textbf{206} & 353 \\
\hline
zeroconf \hspace{2mm} & F[configured] \hspace{2mm} & 268326 & 24 & 413 & \textbf{330} & 376 \\
\hline
\end{longtable}
\end{small}

%% file: 6_related.tex
\section{Related Work}\label{sec:related}

\smallskip\noindent{\bf Strategy representation.}
Previous non-explicit representation of strategies for verification or synthesis purposes typically used BDDs~\cite{WBB+10} or automata~\cite{DBLP:conf/atva/Neider11,DBLP:conf/tacas/NeiderT16} 
and do not explain the decisions by the current valuation of variables.
Classical \emph{decision trees} have been used a lot in the area of machine learning as a classifier that naturally explains a decision \cite{Mitchell1997}.
They have also been considered 
for representation of values and thus implicitly {strategies} for MDP~in \cite{DBLP:conf/ijcai/BoutilierDG95,DBLP:conf/icml/BoutilierD96}.
In the context of verification, this approach has been modified to capture strategies guaranteed to be $\varepsilon$-optimal, for MDPs~\cite{DBLP:conf/cav/BrazdilCCFK15}, 
partially observable MDPs~\cite{BCCGN16}, and (non-stochastic) games \cite{DBLP:conf/tacas/BrazdilCKT18}.
Learning a compact decision tree representation of an MDP strategy was also investigated in~\cite{LPRT10} for the case of body sensor networks.

\smallskip\noindent{\bf Linear extensions of decision trees} have been considered already in~\cite{Dobkin76}
for combinatoric optimization problems. In the field of machine learning, combinations of decision trees and linear models have been
proposed as interpretable models for classification and regression~\cite{breiman84,Quinlan92,frank1998using,landwehr2003logistic}.
A common feature of these works is that they do not aim at classifying the training set without any errors, as in classification tasks this
would bear the risk of overfitting. In contrast, our usage requires to learn the trees so that they fully fit the data.


The closest to our approach is the work of Neider et al.~\cite{DBLP:conf/tacas/NeiderSM16}, which learns decision trees with linear classifiers in the leaves in order to capture functions with generally non-Boolean co-domains.
Since the aim is not to classify, but represent fully a function, our approach is better tailored to representing strategies.
Indeed, since the trees and the lines in the leaves of \cite{DBLP:conf/tacas/NeiderSM16} are generated from counterexamples in the learning process, the following issues arise.
Firstly, each counterexample has to be captured exactly using a generated line.
With the geometric intuition, each point has to lie on a line, while in our approach we only need to separate positive and negative points by lines, clearly requiring less lines.
Secondly, the generation of lines is done online and based on the single discussed point (counterexample). 
As a result, lines that would work for more points are not preferred, while our approach maximizes the utility of a generated line with respect to the complete data set and thus generally prefers smaller solutions.
Unfortunately, even after discussing with the authors of \cite{DBLP:conf/tacas/NeiderSM16} there is no compilable version of their implementation at the time of writing and no experimental confirmation of the above observations could be obtained.

%% file: 7_conclusion.tex
\section{Conclusion and Future Work}\label{sec:concl}
In this work, we consider strategy representation by an extension of decision trees.
Namely, we consider linear classifiers as the leaf nodes of decision trees.
We note that the decision-tree framework proposed in this work is more general.
Consider an arbitrary data structure $\mathscr{D}$,
with an efficient decision oracle for existence of an instance of $\mathscr{D}$ representing
a given dataset without error. Then, our scheme provides a straightforward
way of constructing decision trees with instances of $\mathscr{D}$ as the leaf nodes.

Besides representation algorithms that provably represent entire input strategy,
one can consider models where an error may occur and the data structure is refined
into a more precise one only when the represented strategy is not winning.
Here we can consider more expressive models in the leaves, too.
This could capture representation of controllers exhibiting
more complicated functions, e.g. quadratic polynomial capturing that a robot
navigates closely (in Euclidean distance) to a given point, or deep neural networks
capturing more complicated structure difficult to access directly~\cite{KontschiederFCB16}.

\smallskip\noindent{\bf Acknowledgments.}
This work has been partially supported by
DFG Grant No KR 4890/2-1 (SUV: Statistical Unbounded Verification), 
TUM IGSSE Grant 10.06 (PARSEC), 
Czech Science Foundation grant No. 18-11193S, 
Vienna Science and Technology Fund (WWTF) Project ICT15-003, 
the Austrian Science Fund (FWF) NFN Grants S11407-N23~(RiSE/SHiNE) 
and S11402-N23~(RiSE/SHiNE). 



%% file: 9_appendix.tex
\section*{Appendix}
\appendix


\section{Linear least squares problem}\label{app:linls}

In this work, we consider dataset $\train$ to be a set of tuples (samples) over natural numbers $\vec{x} \in \Nats^d$.
Consider these samples arbitrarily ordered, then $\train$ can be viewed as a matrix $\mat{X} \in \Reals^{n \times d}$,
where $n$ is the number of samples and $d$ is the dimension of the samples.
Let $\mat{X}_i \in \Reals^d$ denote the i-th sample, let $\mat{X}_{i,j} \in \Reals$ denote the j-th dimension in the i-th sample.
We remind that $\train$ is partitioned into good samples $\good$ and bad samples $\bad$.
Consider a vector $\vec{y} \in \set{-1, 1}^n$ constructed as follows: $\vec{y}_i = 1$ iff the i-th sample ($\mat{X}_i$) is $\good$.

The \emph{linear least squares} problem gets a matrix $\mat{X} \in \Reals^{n \times d}$ and a vector $\vec{y} \in \Reals^n$
as input, and outputs a vector of weights $\vec{w} \in \Reals^d$ that minimizes a certain error metric, i.e.,
$\vec{w} = \argmin_{\bar{\vec{w}}} S_{\mat{X}, \vec{y}}(\bar{\vec{w}})$. The error metric to be minimized
in this problem is the \emph{squared error}, formally
\[
S_{\mat{X}, \vec{y}}(\bar{\vec{w}}) = \sum_{i=1}^n \Biggl\lvert\biggl(\sum_{j=1}^p \mat{X}_{i,j}\bar{\vec{w}}_j\biggr) - \vec{y}_i\Biggr\rvert
= \lvert\lvert \mat{X}\bar{\vec{w}} - \vec{y} \rvert\rvert_2^2
\]

When the columns of $\mat{X}$ are linearly independent (which is mostly the case in strategy datasets),
the minimization has a unique solution $\vec{w}$, and the closed-form expression to obtain the solution is as follows.
\[
\vec{w} = (\mat{X}^T\mat{X})^{-1}\mat{X}^T\vec{y}
\]

In case when some columns of $\mat{X}$ are linearly dependent, the solution is no longer unique. To obtain some solution,
one considers an arbitrary maximum subset of linearly independent columns, and then the above expression can be used
with the resulting sub-matrix.

In this work, we use the linear least squares problem to obtain a set of weights for strategy subsets that are not
linearly separable. In such a case, the solution of this problem gives us a classifier (with bias $0$) that, intuitively,
attempts to minimize the separation error on average (following the squared error metric). We obtain such
classifiers in Lines~\ref{line:split_auc_sat}~and~\ref{line:split_auc_unsat} of \cref{alg:split_auc},
and subsequently we compute the area under the ROC curve for them.

\section{Experimental details}\label{app:exp}

Here we provide additional details about the experiments and the results. We have implemented
all algorithms in Python, using the scikit-learn library to manipulate classifiers, ROC curves, etc.
For our experiments we have used a Linux machine with Intel(R) Xeon(R) CPU
E5-1650 v3 @ 3.50GHz (12 CPUs) and 128GB of RAM.

\begin{figure}
\begin{center}
\includegraphics[scale=0.38]{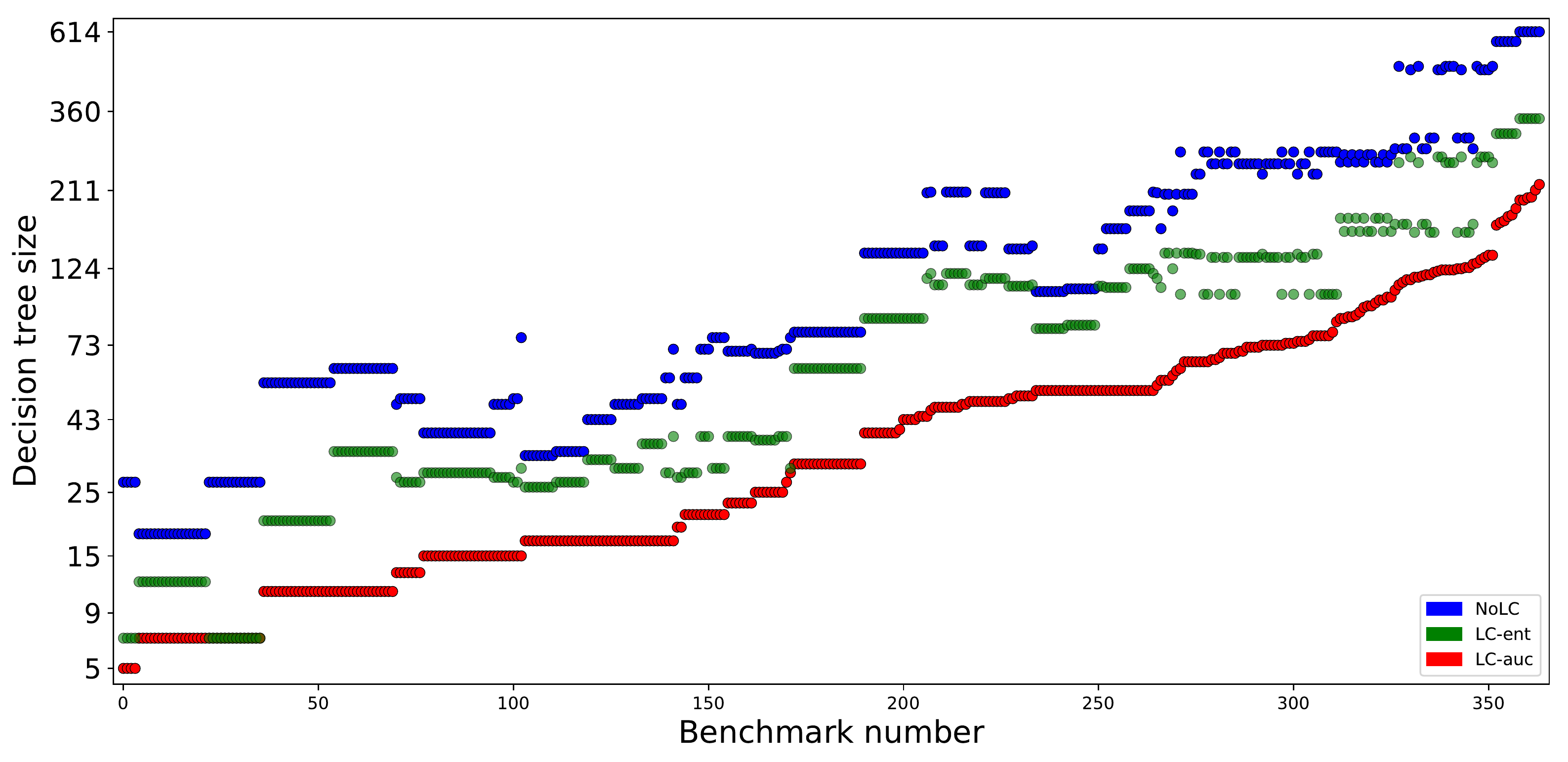}
\end{center}
\caption{Scheduling of Washing Cycles -- overall summary.}
\label{fig:one_wash}
\end{figure}

Considering linear classifiers instead of pure leaves ($\dagger$ instead of $\star$) has typically a
detrimental effect on the construction time. However, this is not necessarily always a case,
as sometimes the decrease in output tree size outweights the overhead of testing linear
separability for classifiers.

Considering the $\auc$ split procedure instead of the information-gain one ($\ddagger$ instead of $\dagger$)
caused a significant increase in construction time. However, we stress that
this is majorly due to our prototypical implementation. In the $\auc$ split procedure, the main loop (Line~\ref{line:split_auc_loop})
considering all possible predicates is embarrassingly parallel. Hence an optimized parallel implementation
of the procedure is expected to suffer minimal overhead in construction time.

\begin{figure}
\begin{center}
\includegraphics[scale=0.38]{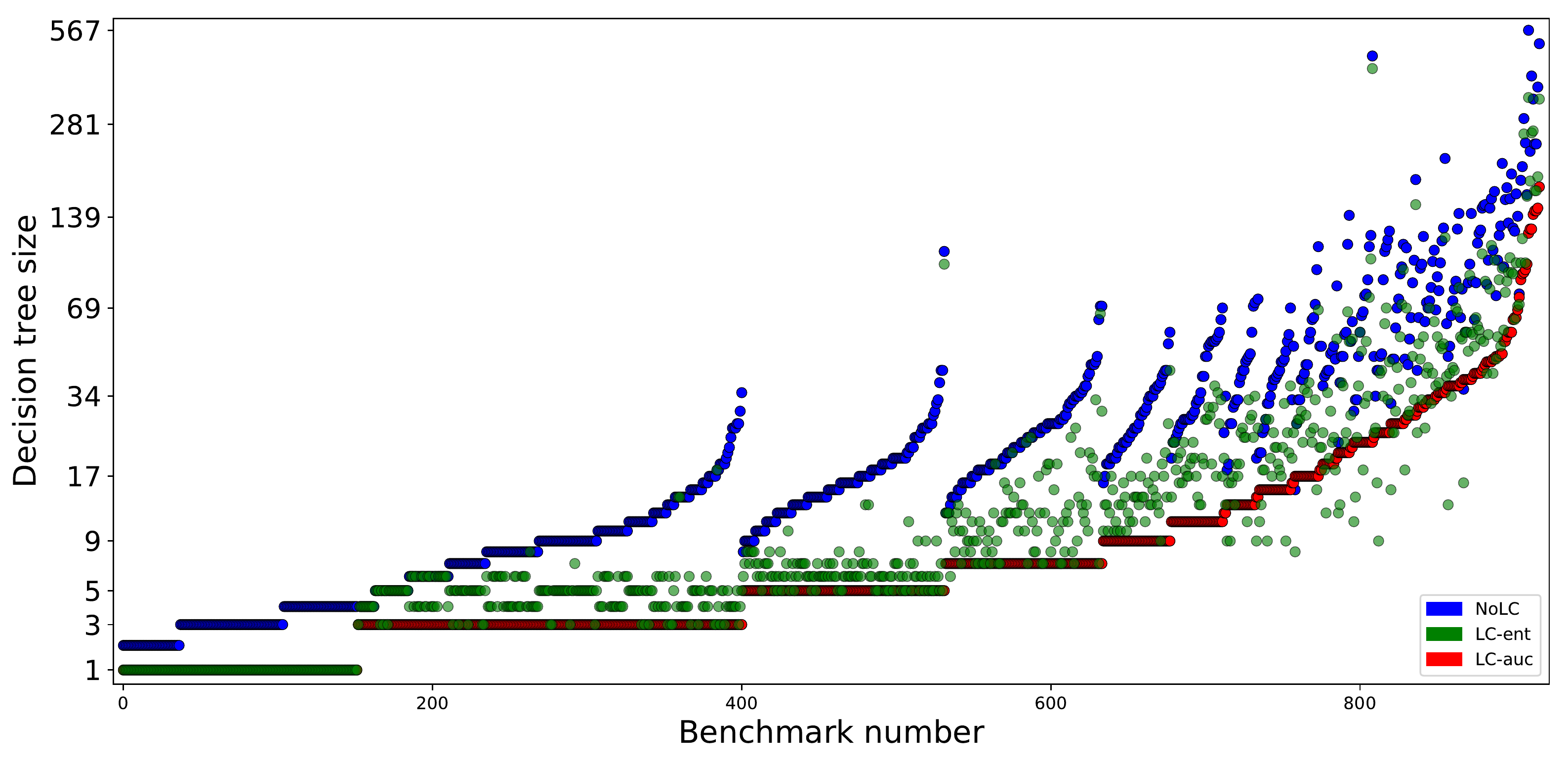}
\end{center}
\caption{LTL synthesis -- overall summary.}
\label{fig:one_ltlgames}
\end{figure}

\begin{figure}
\begin{center}
\includegraphics[scale=0.38]{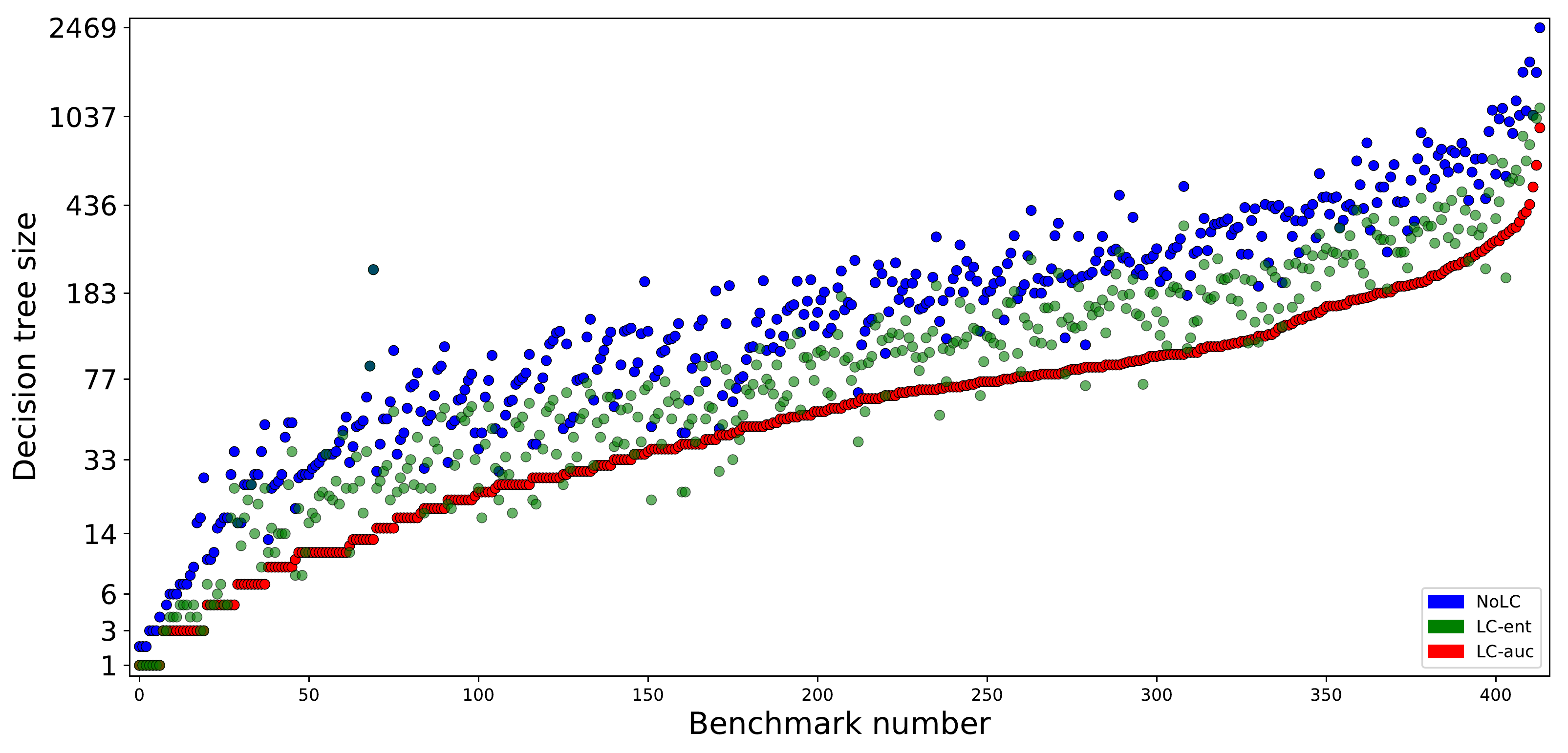}
\end{center}
\caption{LTL synthesis with randomized environment -- overall summary.}
\label{fig:one_ltlmdps}
\end{figure}

\cref{fig:one_wash}, \cref{fig:one_ltlgames}, and \cref{fig:one_ltlmdps} provide a one-plot summary
of the experimental results for the safe scheduling of washing cycles, LTL synthesis, and
LTL synthesis with randomized environment, respectively. In all three figures, each column corresponds
to a benchmark. The colored dots capture the sizes of decision trees for the considered algorithms, namely,
blue for $\star$, green for $\dagger$, and red for $\ddagger$.

